\documentclass{aa}
\usepackage{txfonts}
\usepackage{graphicx}
\begin{document}
%

   \title{The IMF and star formation history of the stellar clusters
   in the Vela D cloud
      \thanks{Based on observations collected at the European Southern Observatory, Chile}}


   \author{F.\ Massi\inst{1} \and
	  L.\ Testi\inst{1} \and
          L.\ Vanzi\inst{2}
          }

   \offprints{F. Massi, \email{fmassi@arcetri.astro.it}}

   \institute{INAF - Osservatorio Astrofisico di Arcetri, Largo E.\ Fermi 5,
	      I-50125 Firenze, Italy
         \and
              ESO, Alonso de Cordova 3107, Vitacura, Casilla 19001, Santiago 19, Chile
             }

   \date{Received ...; accepted ...}

    


\abstract{
We present the results of a Near-Infrared deep photometric survey of a
sample of six embedded star clusters in the Vela-D molecular cloud,
all associated with luminous ($\sim 10^{3}$ L$_{\sun}$) IRAS
sources. The clusters are unlikely to be older than a few $10^{6}$
yrs, since all are still associated with molecular gas.
We employed the fact that all clusters lie at the same distance and
were observed with the same instrumental setting to derive
their properties in a consistent way, being affected by the same
instrumental and observational biases.
We extracted the clusters' $K$ Luminosity Functions and developed a
simple method to correct them for extinction, based on colour-magnitude
diagrams. The reliability of the method has been tested by
constructing synthetic clusters from theoretical tracks for pre-main
sequence stars and a standard Initial Mass Function.
The clusters' Initial Mass Functions have been derived from the
dereddened
$K$ Luminosity Functions by adopting a set of pre-main sequence
evolutionary tracks and assuming coeval star formation.
All clusters are small ($\sim 100$ members) and compact (radius
$\sim 0.1-0.2$ pc); their most massive stars are
intermediate-mass ($\sim 2-10$ M$_{\sun}$) ones. 
The dereddened $K$ Luminosity Functions are likely to arise
from the same
distribution, suggesting that the selected clusters have quite
similar Initial Mass Functions and star formation histories.
The Initial Mass Functions are consistent with those 
derived for field stars and clusters. Adding them
together we found that the ``global'' Initial Mass Function
appears steeper at the high-mass end and
exhibits a drop-off at $\sim 10$ M$_{\sun}$. In fact,
a standard Initial Mass Function would
predict a star with $M > 22.5$ M$_{\sun}$ within one of the clusters,
which is not found. Hence, either high-mass stars need larger
clusters
to be formed, or the Initial Mass Function of the single clusters is 
steeper at the
high-mass end because of the physical conditions in the parental gas.

}
   \titlerunning{Stellar clusters in the Vela~D cloud}

   \maketitle
%

\section{Introduction}

It is now well established by means of direct and indirect observations that
most, if not all, stars are formed in groups rather than in isolation
(Clarke et al.~\cite{CBH00}, Lada
\& Lada 2003). Adams \& Myers (2001) suggested that most stars form in
small clusters with $\sim 10-100$ members. However, by studying
a large sample of embedded clusters within 2 kpc, Lada \& Lada (2003)
concluded that 90 \% of stars that form in embedded clusters occur in rich clusters
of $> 100$ members, but only $< 4-7$ \% of embedded clusters 
emerge from molecular clouds to become bound clusters of Pleiades age.

An important result that strongly
constrains theories of massive stars and stellar clusters formation is that
the stellar density of young stellar clusters seems to depend on the most
massive star in the cluster. In low-mass star forming regions,
stars are usually found to form in
loose groups with typical densities of a few stars per cubic parsec (Gomez et
al.~\cite{Gea93}), while high-mass stars are found in dense clusters of up to
10$^4$ stars per cubic parsec, along with a large fraction of low-mass stars 
(e.g. the Orion Nebula Cluster, Hillenbrand \&
Hartmann~\cite{HH98}). The transition between these two modes of formation
should occur in the intermediate-mass regime, namely 2$\la {\rm
M/M_\odot} \la$15.

In order to probe this transition, Testi et al.~(\cite{TPN99})
carried out
an extensive Near-Infrared (NIR) survey searching for young clusters around
optically visible intermediate-mass stars (Herbig Ae/Be stars) in the northern
hemisphere. The main result of this survey is that there is a strong
correlation between the spectral type of the Herbig Ae/Be stars and the
membership number of the stellar groups around them. Furthermore, there
is compelling evidence that the most massive stars in their sample
are surrounded by {\em denser}, not only more populous, clusters.
These findings are in qualitative agreement with
models that suggest a causal relationship between the birth of a massive
star and the presence of rich stellar clusters. The observed
correlation and scatter, however, could also be explained in terms
of random assembling clusters with membership size distribution of the
form $g(N)\sim N^{-1.7}$ picking stars from a standard Initial Mass Function
(IMF; Bonnell \& Clarke~\cite{BC99}).

As discussed in Testi et al.~(\cite{TPN01}a)
there are two observational strategies that may
provide additional constraints on which of the two scenarios is the most
appropriate: expand the sample of optically revealed young O and B stars to
increase the statistic, and search for clusters in complete samples of
luminous embedded sources in giant molecular clouds. 
Recently, de Wit et al.~(\cite{WJa}; \cite{WJb}), 
following the first approach,
investigated the possible origin in clusters
of field O-stars; they demonstrated that the majority of the so called
``field'' O-stars may have originated in clusters, but could not
exclude that a small fraction of these are actually born in isolation.
An example of the second approach is given in Massi et al. (2000, 2003) 
as described in Sect.~\ref{VMRD}. 

One fundamental property of clusters is their IMF; different methods to
extract the IMF from photometric and/or spectroscopic data are summarized 
in Lada \& Lada (2003). These authors concluded that
embedded young clusters exhibit strong similarities in their IMF
with open clusters and field stars, although this ``universal'' IMF 
does not seem to characterize all star forming regions (Lada \& Lada 2003).
Scalo (1998), reviewing all determinations of the IMF in different
environments, noted
that either the observational uncertainties are so large as 
not to allow one to draw definite conclusions on the average IMF and its variation,
or the available data show strong evidence for IMF variations. 
This is a critical issue, since the IMF should represent the imprint 
on a stellar population of the physical conditions in the parental gas. 

Most of the main topics concerning the formation of star
clusters are then still debated and ask for large observational
efforts.  In this paper, we report the results of a NIR study of 
a sample of small clusters in the giant molecular cloud ``D'' of
the Vela Molecular Ridge. Small young embedded clusters can prove very interesting 
targets to address many of the illustrated points, since they 1) are associated
with the intermediate-mass transition regime 
investigated by Testi et al.~~(\cite{TPN99}),
2) are on the border between the different sizes envisaged by Adams \& Myers (2001) and
Lada \& Lada (2003) as major star contributors, and 3) do not host stars massive enough
to heavily affect the parental environment, maintaining a stronger link with it.
All imaged clusters are young and are associated with the same molecular cloud.
On the one hand, this allows one to circumvent many effects due to 
observational biases in comparing the results. On the other hand, local
differences in the global IMF emerging from the molecular cloud can
be searched for and investigated. Furthermore, our data considerably improve on
previous NIR photometry of the same regions, both in sensitivity and
in field of view.
The Paper layout is as follows: in
Sect.~\ref{VMRD}, the main features of the star forming region and the selected
sample are summarized. Observations and data reduction are described in
Sect.~\ref{obs:s}. The main results are reported in Sect.~\ref{res:s}
and discussed in Sect.\ref{disc:s}. Our conclusions are listed in
Sect.~\ref{conc:s}.

\section{The Vela-D Luminous IRAS Sources}
\label{VMRD}

The Vela Molecular Ridge (VMR) is a giant molecular cloud complex located in
the outer Galaxy ($257 \degr < l <  274 \degr$) within the galactic plane
($-5 \degr < b < 5 \degr$). It was first mapped in the CO(1--0)
transition (with low resolution) by Murphy \& May (1991), who
divided the emission area into
4 main regions (named A, B, C and D) corresponding to local maxima. The 
kinematical distance is 1--2 kpc. 

Liseau et al.\ (1991) and Lorenzetti et al.\ (1993) 
studied
the star formation in the VMR by selecting the IRAS point sources with red
colours and combining their Far-Infrared (FIR) fluxes 
with NIR and 1.3-mm photometry.
They list a sample of IRAS sources which, based on the SEDs from NIR to
FIR wavelengths and the radial
velocity of the associated molecular gas, may be considered 
as intermediate-mass ($M \sim 2-10$ M$_{\sun}$) analogues of Class I sources 
belonging to the VMR. Liseau et al.\ (1991) also
discuss the issue of distance, concluding that clouds A, C and D are at
$d= 700 \pm 200$ pc, whereas cloud B is further out ($\sim 2$ kpc). 
Massi et al.\
(2000) analyzed $JHK$ images of the fields towards 12 sources 
of the sample, namely those associated with cloud D, whose bolometric 
luminosities range from $\sim 10^{2}$ to $\sim 6 \times 10^{3}$ 
L$_{\sun}$ (Massi et al.\ 1999). 
These authors found young embedded star clusters towards the IRAS
sources with $L_{\rm bol} \ga  10^{3}$ L$_{\sun}$. 
Five of these embedded clusters have been selected for the present work.
In addition, we have included IRAS 08438--4340 
(IRS 16, as designated by Liseau et
al.\ 1991), a FIR source associated with both an embedded 
star cluster (Massi et al.\ 2003) and an HII region,
close (in projection) to the other ones. The six fields are listed in
Table~\ref{source:list} along with their equatorial coordinates and 
the nomenclature used by Liseau et al.\ (1991).

%
\begin{table}
\caption[]{Selected fields. The last Col.\ lists the designation
 adopted by Liseau et al.\ (1991). \label{source:list}}
\begin{tabular}{cccccccc}\hline
\\
IRAS source & \multicolumn{3}{c}{$\alpha$(1950.0)}& 
\multicolumn{3}{c}{$\delta$(1950.0)}& \\
& $^{h}$ & $^{m}$ & $^{s}$ & $^{\degr}$ & $^{\arcmin}$ & $^{\arcsec}$& \\
\hline
IRAS 08438--4340 & 08 & 43 & 50.2 & -43 & 40 & 02 & IRS 16 \\
IRAS 08448--4343 & 08 & 44 & 49.4 & -43 & 43 & 27 & IRS 17 \\
IRAS 08470--4243 & 08 & 47 & 00.0 & -42 & 43 & 12 & IRS 18 \\
IRAS 08470--4321 & 08 & 47 & 01.3 & -43 & 21 & 15 & IRS 19 \\
IRAS 08476--4306 & 08 & 47 & 39.4 & -43 & 06 & 01 & IRS 20 \\
IRAS 08477--4359 & 08 & 47 & 47.1 & -43 & 59 & 34 & IRS 21 \\
\hline
\end{tabular}
\end{table}
%
%
%

\section{Observations and data reduction}
\label{obs:s}

The six fields towards the selected IRAS sources
in the Vela D molecular cloud were observed at the ESO-NTT telescope
on February 18, March 5, March 20 and September 25, 2000.
Near-Infrared images through the standard $J$, $H$, and $K_{s}$ 
broad band filters were acquired using the SofI instrument, with 
a plate scale of $\sim$0.282\arcsec/pixel and an instantaneous field of 
view of $\sim$5$^{\prime}$. For each target we collected a set of
dithered exposures for a total integration time of 15 minutes in each filter.
The raw images were crosstalk corrected, flat fielded, sky subtracted, aligned
and mosaiced using the special procedures developed for SofI and standard
routines within the IRAF package.

The final image quality of our images was $\sim 0.85$\arcsec\ for all but the 
field surrounding IRS~16 which was observed under slightly worst seeing 
conditions (1.1\arcsec). Photometric calibration was ensured by observations 
of a number of standard stars from the list of Persson et al.~(\cite{Pea98}).
Source detection and aperture photometry were performed using standard routines 
within
IRAF. 

We compared our photometry with that of Massi et al.\ (1999) and Massi et al.\
(2003) to check for consistency. The latter was obtained from Irac2 $JHK$ images.
The mean differences (in the sense Irac2 minus SofI) are: $\sim 0$ to $-0.07$ mag
(excepted IRS 19 and IRS 21, with $-0.34$ and $-0.16$ mag) in $K$, 
$-0.06$ to $0.01$ mag (excepted IRS 17 and IRS 21, with $0.3$ and $-0.2$ mag) in
$H$
and $-0.1$ to $0.08$ mag (excepted IRS 17 and IRS 16, with $0.26$ and $-0.16$
mag) in $J$. Part of the largest mean differences may be attributed to problems with the
aperture correction of some Irac2 data sets, as already noted by those authors
(Massi et al.\ 2000; see also their Fig.~1 for IRS 17 and IRS 19).
The scatter of the differences around the average values are
$0.11$ to $0.26$ mag in $K$, $0.12$ to $0.19$ mag (excepted IRS 16 and IRS 17,
with $0.34$ and $0.33$ mag) in $H$ and $0.21$ to $0.25$ mag (excepted IRS 16 and IRS
17, with $0.36$ and $0.30$ mag) in $J$.
We found also a colour term in $K$ of the order of $\sim -0.2 \times (H - K)$,
certainly due to the different $K$ (Irac2) and $K_{s}$
(SofI) filters, that may account for another part of average and r.m.s.
differences. In $H$ and $J$ the colour term, if any, is much smaller:
$\sim 0.04-0.07 \times (J - H)$. We believe that the relatively 
large scatter in the magnitude differences may be explained by the differences
in PSF sampling and sky estimate between the data sets and the intrinsic
variability of young stars. Anyway, both data sets appear to be mutually consistent.

\section{Results}
\label{res:s}

%
%
\begin{figure*}
\centerline{\includegraphics[width=16cm]{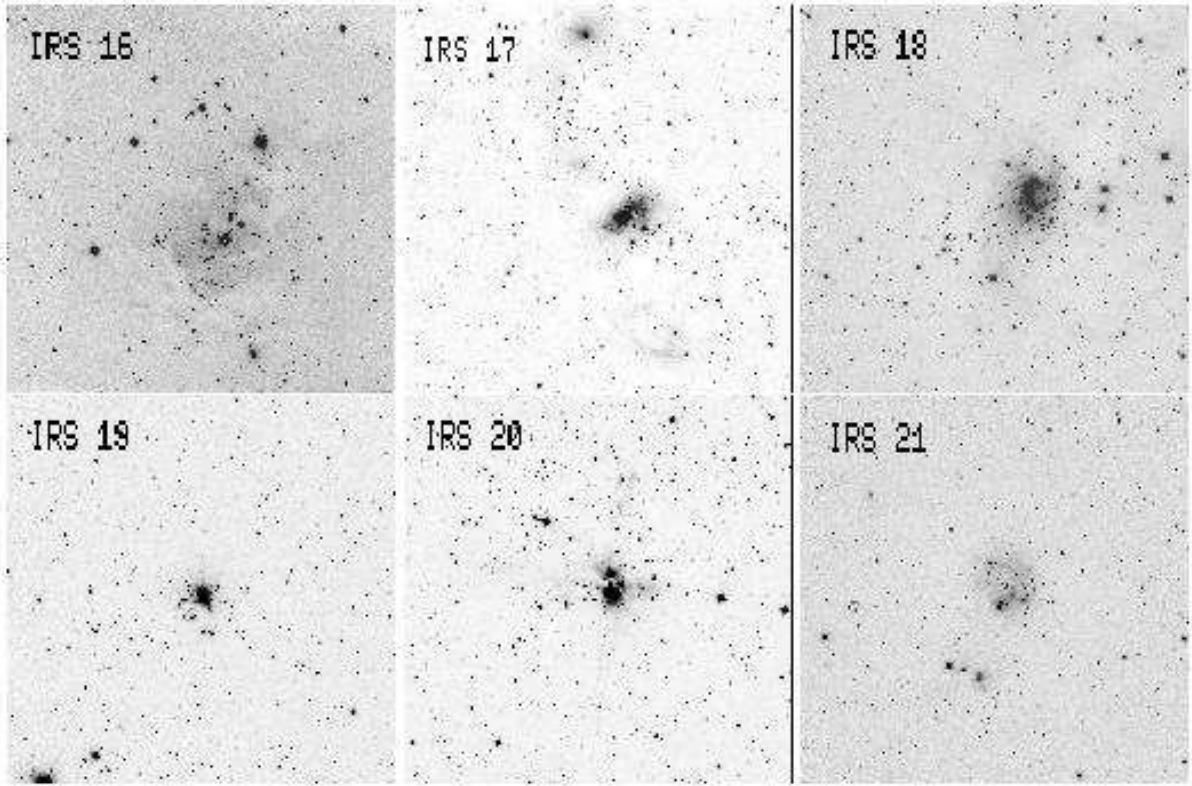}}
\caption[]{NTT/SofI Near-Infrared images (through the $K_{s}$
band) of the fields centred on the selected luminous IRAS sources 
associated with the Vela~D molecular
cloud.} 
\label{ftruecol}
\end{figure*}
%

In Fig.~\ref{ftruecol} we show the NTT/SofI $K_{s}$ 
images of the six surveyed fields.
True-colour images (combining the images in all three bands) are
shown in Testi et al.\ (2001b; see their Fig.~1).
From those images it is immediately clear that we detected groups of very red 
sources in every fields. 

\subsection{Mass sensitivity}
We estimate a completeness magnitude $K^{\rm c}_{s} \sim 18$ for all
fields, excepted IRS 21 for which $K^{\rm c}_{s} \sim 17$
(see Fig.~2 in Testi et al.\ 2001b). Assuming a distance
modulus of 9.22 mag, this allows an estimate of the minimum mass, as a function of age
and reddening, for a star whose probability to be found is still almost 1 throughout each
$K_{s}$ frame. Using the evolutionary tracks for pms stars given by Palla \&
Stahler (1999), the mass completeness limit is $M_{\rm compl} \sim 0.1$ M$_{\sun}$ 
for $A_{V} = 30$ mag, $K^{\rm c}_{s} \sim 17$ and an age of $10^{6}$ yrs and 
increases to $\sim 0.4$ M$_{\sun}$ for
an age of $10^{7}$ yrs. At this age, for $A_{V} = 20$ mag or $K^{\rm c}_{s} = 18$, 
$M_{\rm compl}$ is still as low as 0.2 M$_{\sun}$. We find the same using the
evolutionary tracks from D'Antona \& Mazzitelli (1994) for $10^{6}$ yrs old stars, but
$M_{\rm compl}$ ranges from $\sim 0.5$ M$_{\sun}$ ($A_{V} = 30$ mag) to
$\sim 0.3$  M$_{\sun}$ ($A_{V} = 20$ mag or $K^{\rm c}_{s} = 18$) for
$10^{7}$ yrs old stars. From Baraffe et al.\ (1998) the estimated $M_{\rm compl}$ is
$\sim 0.15$ M$_{\sun}$ for $2 \times 10^{6}$ yrs old stars and $\ga 0.6$ 
M$_{\sun}$ for $10^{7}$ yrs old stars. It is clear that our images probe the 
clusters stellar population down
to the hydrogen burning limit for ages $< 10^{6}$ yrs and as low as $0.5$
M$_{\sun}$ for an older population ($10^{7}$ yrs); they are probably deep enough
to probe the brown dwarf regime for ages $\sim 10^{5}$ yrs. However, we caution
against the decrease in completeness magnitude that is likely to occur towards the
clusters' centres because of crowdedness, diffuse emission and, sometimes, the
PSF wings of the brightest stars.

\subsection{Clustering}
\begin{figure*}
\centerline{\includegraphics[width=16cm]{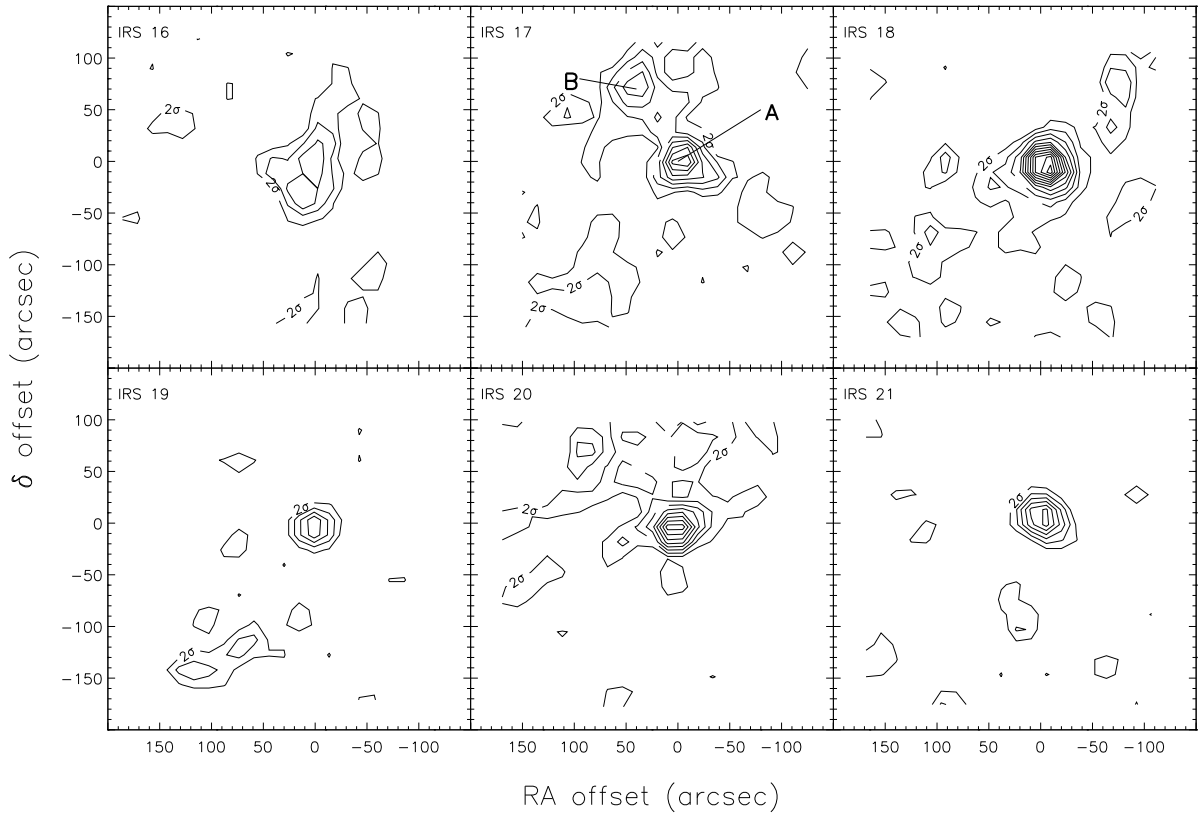}}
\caption[]{Star surface density (arcmin$^{-2}$) from the $K_{s}$ images
of the 6 fields. Contours are in steps of $\sim 2\sigma$ (20 arcmin$^{-2}$)
from $\sim 2\sigma$ above the mean surface density
of field stars. Towards IRS 17, two nearby subclusters (A and B) are clearly
present.
\label{densmap}}
\end{figure*}

Although the star clusters are clearly visible in the images of Fig.~\ref{ftruecol},
we employ the same graphical representation as used by Massi et al.\ (2000) in 
order to better evidence their presence. Hence, we counted all $K_{s}$ sources up
to $K_{s} = 18$ in $\sim 30\arcsec \times 30 \arcsec$ squares (``bins''), 
at offsets of
$\sim 15\arcsec$ both in right ascension and in declination. Following Massi
et al. (2000), we estimated the field star density (the ``sky'') and its
$1 \sigma$ fluctuation by gathering all the bins from the 6 fields and constructing
a global histogram of the number of bins with given counts. 
The rising part of the histogram
is well fitted by a Poissonian curve with a mean of $7.5$ counts per bin (32 stars
arcmin$^{-2}$), so we assume that this represents the statistic of field stars.
The actual data distribution exceeds the Poissonian curve in the wing and, as noted by
Massi et al. (2000), this indicates that the clustering is not due to sky
fluctuations. The Poissonian fit yields a $\sigma  \sim 12$ stars arcmin$^{-2}$.
Since our imaged fields are large enough, we constructed a histogram for each
frame and by the same technique we obtained sky means ranging from 26 to 34
stars arcmin$^{-2}$ (and $\sigma \sim 10-12$ stars arcmin$^{-2}$). These values
compare quite well with those found by Massi et al. (2000), i.\ e., a sky mean
$\sim 20$ stars arcmin$^{-2}$ and a $\sigma \sim 12$ stars arcmin$^{-2}$, 
considering that their photometry has a limiting magnitude $K \sim 18$ 
and, hence, a lower completeness magnitude than ours. Figure~\ref{densmap}
shows the contour maps of surface star density from the $K_{s}$ counting;
contours start at 50 stars arcmin$^{-2}$ (roughly the sky mean plus $2 \sigma$) in steps of
20 stars arcmin$^{-2}$ ($\sim 2$ $\sigma$). Clearly, all clusters have maximum
densities well above the sky mean plus $3 \sigma$. This confirms that the
structures found by Massi et al.\ (2000) on smaller field-size images are actually
young star clusters. The clusters are compact enough to fit the $2 \arcmin \times 
2 \arcmin$ f.o.v. of the images used by those authors. However, the larger f.o.v.
allowed by SofI and our much deeper imaging
evidence some substructuring, such as the small source grouping 
(which we name ``B'') north-east of the cluster found by Massi et al.\ towards IRS 17
(which we name ``A'').
%
%
%
%
\begin{table}
\caption[]{Clusters' properties. The diameter is
listed in Col.\ 2, the peak surface stellar density in Col.\ 3,
the richness indicator (see text) in Col.\ 4 and
the volume stellar density in Col.\ 5. \label{clust:prop}}
\begin{tabular}{ccccc}\hline
Field & Size  & Max.\ surf.\ & I$_{\rm c}$ & Vol. \\
name  &    & density &             & density \\
      & (pc) & (arcmin$^{-2}$) &    & (pc$^{-3}$) \\
\hline
IRS 16 & 0.26 & 100 & $71 \pm 11$ & 1830 \\
IRS 17A & 0.21 & 180 & $50 \pm 8$ & 1289 \\
IRS 17B & 0.22 & 120 &  & \\
IRS 18 & 0.27 & 270 & $118 \pm 13$ & 3042\\
IRS 19 & 0.13 & 120 & $35 \pm 8$ & 902 \\
IRS 20 & 0.21 & 190 & $59 \pm 8$ & 1521 \\
IRS 21 & 0.16 & 150 & $52 \pm 8$ & 1340 \\
\hline
\end{tabular}
\end{table}
%
%
%

\begin{figure*}
\centerline{\includegraphics[width=12cm]{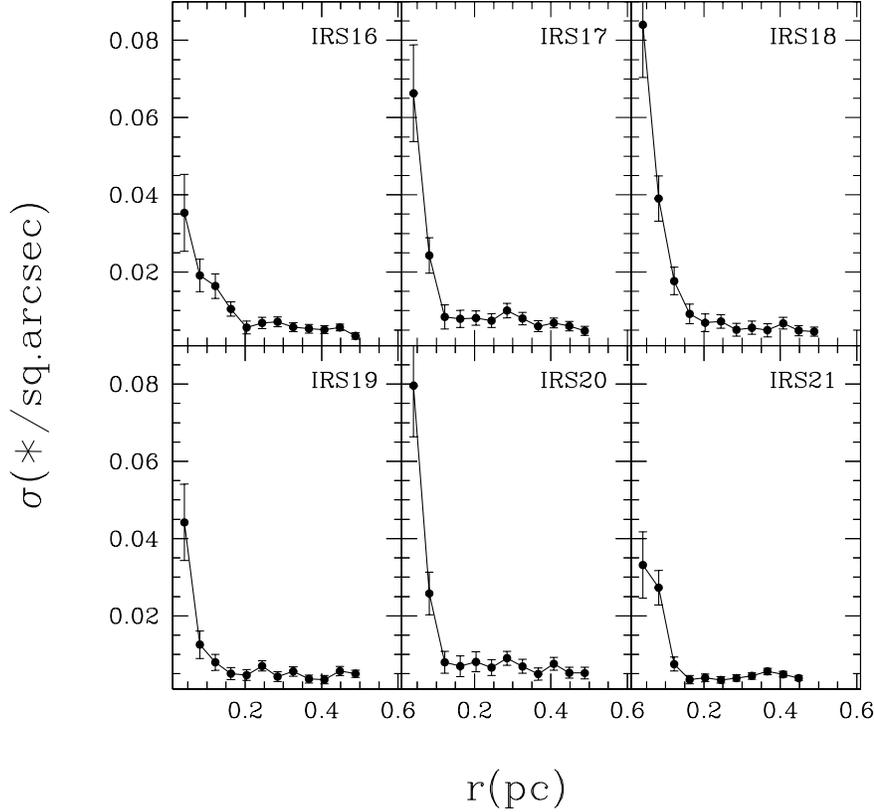}}
\caption[]{Radial distribution of star surface density (arcmin$^{-2}$) 
from the $K_{s}$ images of the 6 fields. 
\label{rad:dens}}
\end{figure*}

We determined the radial distribution of surface star density as described in
Testi et al.\ (1999), shown in Fig.~\ref{rad:dens}. Counting is 
performed on the sources with at least a detection in $K_{s}$ and up to $K_{s}= 17$,
which is the smaller completeness magnitude we obtained throughout the fields.
A significant density
increase towards $r=0$ pc is noticeable in all the six fields. 
This also allows an estimate of the surface density due to field stars by 
averaging the radial density in the wing of the distribution. Obtained
values range between 14 and 25 stars/arcmin$^{2}$ and are consistent with
the one of 32 stars/arcmin$^{2}$ we infer above, considering the latter 
is obtained for sources up to $K_{s} = 18$. The richness indicator
I$_{\rm C}$ (Testi et al.\ 1999) is derived by integrating
the radial stellar surface density, once subtracted the background-foreground
contribution as estimated from the wing of the distribution.
This is listed in Table~\ref{clust:prop} along with
the statistical error due to the sky determination and the Poissonian fluctuations
in the surface density of the cluster members. Figure~\ref{rad:dens} shows that
all clusters have a rather similar size and allows to estimate a radius $\sim
0.1-0.2$ pc ($\sim 0.5-1$ arcmin) at the assumed distance (see Table~\ref{clust:prop}).
%
%
%

\subsection{Cluster star population}
\label{csp}
\begin{figure*}
\centerline{\includegraphics[width=16cm]{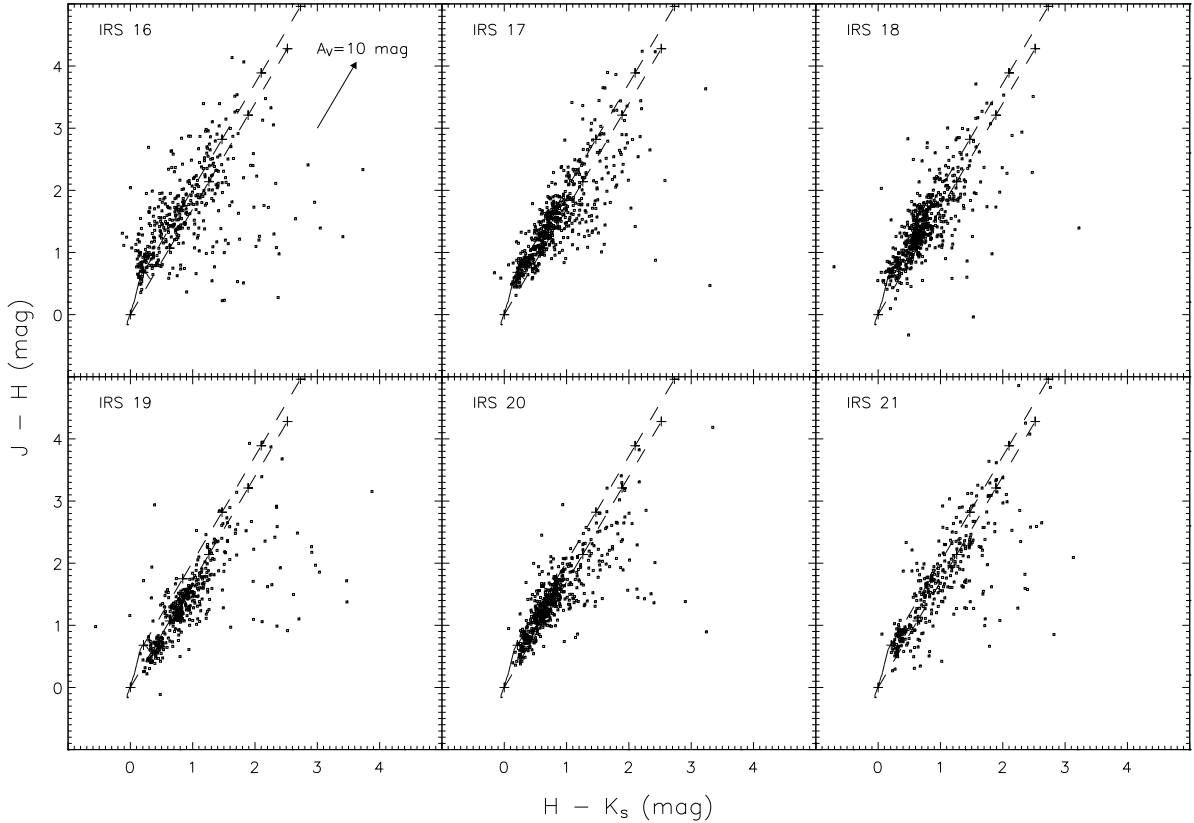}}
\caption[]{Colour-colour diagrams (i.\ e., $J - H$ vs.\ $H - K_{s}$) of
the 6 fields, for objects lying within 2 arcmin (i.\ e., roughly
twice the clusters' radius) from the clusters' centres
and up to $K_{s} = 18$. The solid line marks the main sequence
(Koornneef 1983) whereas the dashed ones indicate the reddening law with
crosses at intervals of $A_{\rm V} = 10$ mag (Rieke \& Lebofsky 1985).
For the sake of clarity, an arrow corresponding to an extinction
$A_{\rm V} = 10$ mag has been also drawn, in the upper left panel.
\label{col-col-in}}
\end{figure*}

\begin{figure*}
\centerline{\includegraphics[width=16cm]{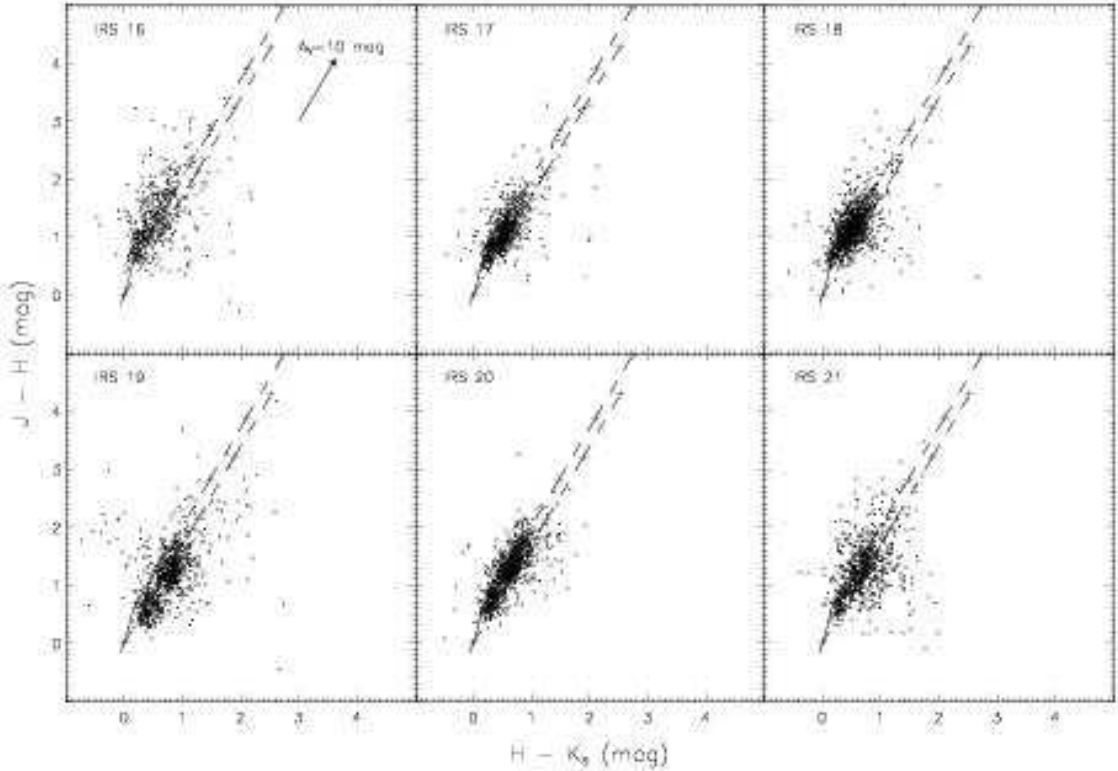}}
\caption[]{Same as Fig~\protect\ref{col-col-in}, but for objects
 further than 2 arcmin from the clusters' centres.
\label{col-col-out}}
\end{figure*}

The nature of the star population associated with the embedded clusters
can be studied through colour-colour ($J - H$ vs.\ $H - K_{s}$) and
colour-magnitude ($K_{s}$ vs.\ $H - K_{s}$) diagrams. Exploiting 
our field of view, we compare the colours of sources within 2 arcmin of the cluster
centre with those of sources outside. This distance is slightly more than twice
the clusters' radius; in fact, we expect the radial profiles to exhibit a lower density
tail far from the centre, so we chose a value of 2 arcmin in order to account for
this. Radial density profiles on a much larger scale than our images,
obtained from 2MASS data, actually confirm the presence of 
a tail up to at least twice the clusters' radius. Hence, the sky area outside
this distance is less contaminated by clusters' members and more representative of
the population of field stars.  Figure~\ref{col-col-in} shows the
colour-colour diagrams (CCDs) for objects within the chosen radius, whereas
Fig.~\ref{col-col-out} shows the same but for objects outside, both up to
$K_{s} = 18$. The locus of main sequence (from Koornneef 1983) is
drawn as a solid line and the dashed lines are parallel to the reddening
vector (with crosses at intervals of $A_{\rm V} =10$ mag) according to the
extinction law given by Rieke \& Lebofsky (1985). The NIR sources 
$> 2$ arcmin far from the clusters' centres are essentially reddened stars, up to
$A_{\rm V} \sim 10 - 15$ mag; the dispersion around the main sequence and its
reddening band denotes a large fraction of faint objects that may
represent background field stars. There are few NIR sources with an 
intrinsic colour excess.  Conversely, within the chosen radius, 
a different stellar population emerges from the previous one; 
the objects now spread over a larger extinction range,
up to $A_{\rm V} \sim 20 - 30$ mag, there is less dispersion around the main
sequence and the fraction of NIR sources with an intrinsic colour excess is
much larger. This indicates that the peak extinction is greater towards the
clusters' centres (hence, most background stars are reddened out) and that a
population of very young stars is superposed to that of field stars which
characterizes the outer edge of the imaged fields. Including only 
sources with $K_{s} \leq 15$ further limits the dispersion around the main sequence
and reddening bands, but does not alter the spread along the extinction 
vector and the high fraction of NIR objects with intrinsic colour excess near
to the clusters' centre. Also, note that the observed reddening
follows quite well the drawn loci, confirming that the extinction law
towards the 6 fields is consistent with the standard one as given by Rieke \&
Lebofsky (1985).
%
%
%

%
%
%

\begin{figure*}
\centerline{\includegraphics[width=16cm]{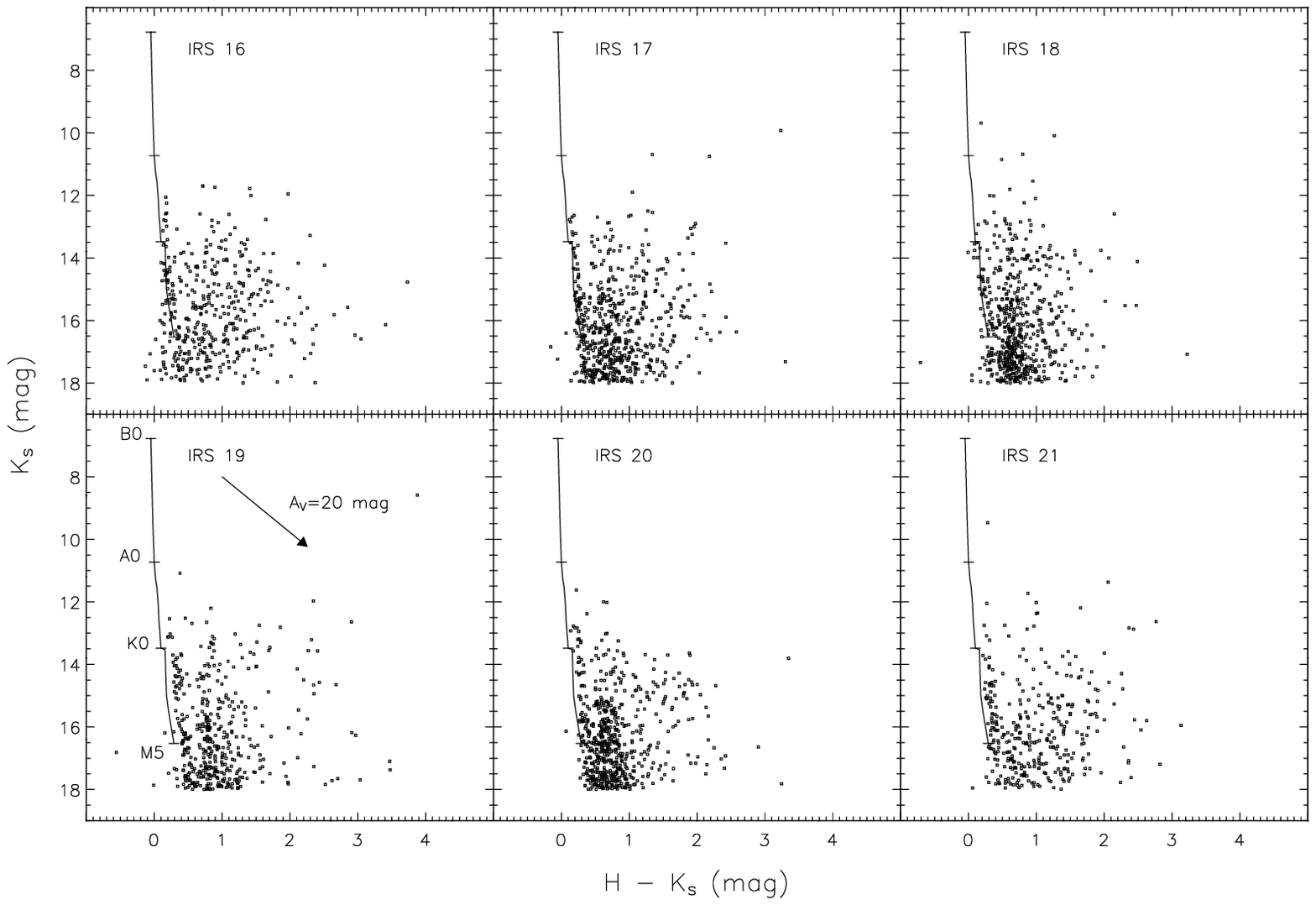}}
\caption[]{Colour-magnitude diagrams (i.\ e., $K_{s}$ vs.\ $H - K_{s}$) of
the 6 fields, for objects lying within 2 arcmin from the clusters' centres
and up to $K_{s} = 18$. The solid line is the Zero Age Main Sequence 
(from Allen 1976 and Koornneef 1983) at the distance of the VMR,
whereas the arrow is for a reddening 
$A_{\rm V} = 20$ mag (Rieke \& Lebofsky 1985). The location of B0, A0,M0 and M5
ZAMS stars is labelled.
\label{mag-col-in}}
\end{figure*}

\begin{figure*}
\centerline{\includegraphics[width=16cm]{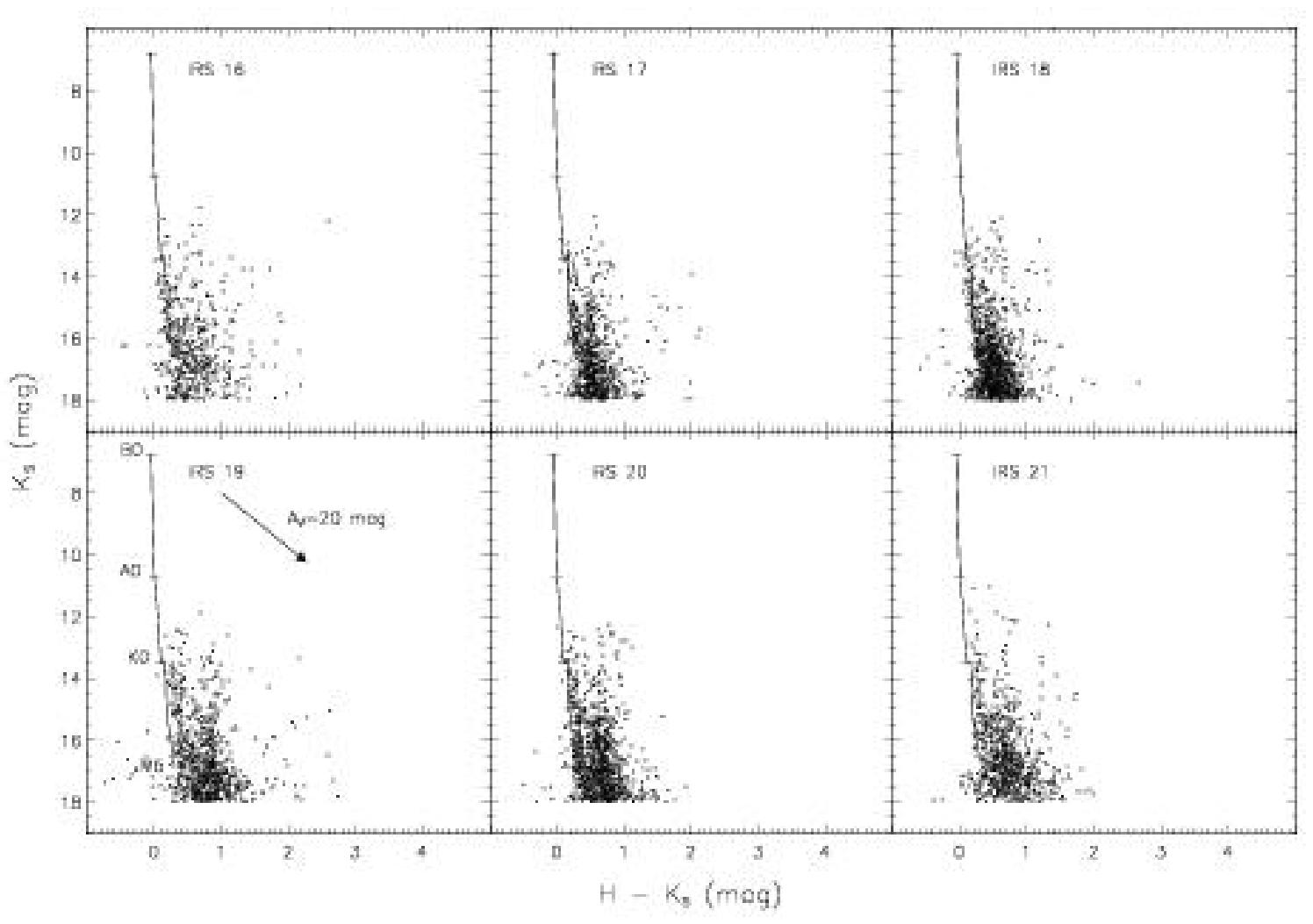}}
\caption[]{Same as Fig~\protect\ref{mag-col-in}, but for objects
 more than 2 arcmin from the clusters' centres.
\label{mag-col-out}}
\end{figure*}

All the previous findings can be checked using colour-magnitude diagrams
(CMDs). These are
presented in Fig.~\ref{mag-col-in} and Fig.~\ref{mag-col-out} for all 
fields. The solid line is the locus of Zero Age Main Sequence (ZAMS) at the
distance of Vela (700 pc) using the absolute magnitudes given by Allen (1976)
and the colours given by Koornneef (1983). The location of B0, A0, K0 and M5 stars
are labelled. The arrow is for a reddening $A_{\rm V}=20$ mag according to
Rieke \& Lebofsky (1985).
It can be
seen that the sources far from the clusters' centres (see Fig.~\ref{mag-col-out})
are on average closer to the ZAMS (i.\ e., less extincted)
and a large fraction of them is fainter
than $K_{s} \sim 16$ (i.\ e., they are mostly
background field stars as noted above). 
Actually, the fraction of sources (up to $K_{s} = 18$)
with $K_{s} < 16$ within/outside 2 arcmin from the centres 
ranges from 0.46/0.33 (IRS 16) to 0.36/0.29 (IRS 20).
Furthermore, the sources within 2 arcmin from the
clusters' centres are in a larger fraction far from the ZAMS, because of both
their greater extinction and their intrinsic colour excess:  
the fraction of sources (up to $K_{s}
=18$) with $H - K_{s} > 1.5$ mag within/outside 2 arcmin from the centres
ranges from 0.18/0.03 (IRS 16) to 0.07/0.007 (IRS 18).
This confirms
that the clusters' stellar population is represented by very young, embedded
stars, mostly located within 2 arcmin (or even less) from the centres and that
it is indeed different from the one of field stars.
%
%
%

%
%
%

\subsection{The clusters' $K$ Luminosity Functions}

The $K$ Luminosity Function (KLF) of an embedded cluster is a simple tool
to probe its Initial Mass Function (IMF) and age. As discussed, e.\ g.,
by Zinnecker et al.\ (1993),
the KLF depends not only on the underlying IMF, but also on the mass-luminosity
relation. For pms stars the mass-luminosity relation is a function of age.
Hence,
given the IMF, a cluster's KLF evolves with time and also changes according to the
star formation history of the cluster. However, Muench et al.\ (2000) show that
the KLF of a young cluster is much more sensitive to variations in its 
underlying IMF than to the star formation history or the mass-luminosity
relation. Also, a cluster KLF shifts to fainter magnitudes and widen systematically with age.
This should allow to put some constraints on the age of a young clusters,
as well.

The KLFs obtained from our $K_{\rm s}$ photometry are shown in Fig.~2 of Testi et
al.\ (2001b). For each field, two KLFs are determined: one using sources close to
the cluster centre and one using sources at the image edge. Clearly,
the two sets of distributions differ in shape, confirming that the clusters  
host a star population different from
the field one, as inferred in Sect.~\ref{csp}
on the basis of CCDs and CMDs. 
In particular, the outer KLFs increase
steadily and peak at higher magnitudes than the ``central'' KLFs. This maximum 
reflects the completeness limit at $K_{s}$. 
Since at the edges of our images the
number of cluster members is small with respect to field stars, 
we chose to construct ``control'' KLFs using the 
photometry obtained in the outer areas themselves of our frames. 
These should better characterize the contribution of field 
stars to the raw KLFs than those obtained using relatively distant control
regions. 
We adopted a radius of $2\arcmin$ from the cluster centre to
define the border between ``centre'' and ``edge''. 
The cluster's centre is assumed to coincide with the NIR counterparts of the IRAS sources 
identified by Massi et al.\ (1999). The reason for our choice of the radius
is explained in
Sect.~\ref{csp}; actually a larger radius would assure a much less contamination, but
would also decrease the outer area available for estimating the control KLF. 

Thus, we constructed the two KLFs per field by counting the $K_{s}$ sources
in bins of 0.5 mag, within and outside the selected radius.
After normalizing the outer KLFs to the same area of the central
ones, the clusters' KLFs were finally obtained by 
subtracting the former from the latter in each field. 
Yet, in order to derive the clusters' IMFs from their KLFs,
source extinction and NIR excess have to be accounted for. 
Extinction causes a twofold problem. 
First, even with so nearby control fields the sources 
at the centre and those at the edge of each image are reddened in different ways,
as evidenced by the CCDs. 
Furthermore, the single clusters' members 
are differentially reddened and, because the span in $A_{V}$ may be more than 
10 mag (i.\ e.,
more than 1 mag in $A_{K}$), this cannot be neglected even binning the KLFs at a
large interval. Thus, we tested two methods in order
to account for the effect of extinction on the KLFs: 
modelling the reddening of theoretical KLFs, and trying to ``deredden'' 
the observed KLFs. We found the latter method more satisfactory and we
discuss it in the following.

\subsection{Modelling $K$-Luminosity functions}

We performed several tests with synthetic KLFs, which are briefly
discussed in the following sections. To do that, 
we developed a Montecarlo procedure to construct theoretical
KLFs by assuming an underlying IMF, an 
age and a star formation history. The adopted mass-luminosity relation has been 
derived from the evolutionary tracks of Palla \& Stahler (1999). 
The procedure makes an
estimate of the number of cluster members on the basis of
the bolometric luminosities that is selected, using an empirical relation 
between I$_{\rm c}$ and $L_{\rm bol}$ derived from the data of Testi et al.\ (1998,
1999). The NIR colour excess has been modelled as in Hillenbrand \& Carpenter 
(2000) and it is also possible to add to the NIR fluxes a random reddening (according
to a Gaussian statistic) superposed to a constant value.

\subsection{Dereddening the observed $K$ Luminosity Functions}

There are indications that the surveyed clusters are at most a few $10^{6}$ yrs old:
their compactness, the high spread in extinction displayed by their members
and a significant fraction of sources
exhibiting a NIR infrared excess in the CCDs 
(see Fig.~\ref{col-col-in}). According to Lada \& Lada (2003),
the embedded phase of cluster evolution lasts $2-3$ Myrs and clusters
older than 5 Myrs are rarely associated with molecular gas. All 
our observed clusters are associated with molecular gas (see Massi et al.\ 2000,
2003, 2005).

\begin{figure}
\centerline{\includegraphics[width=8cm]{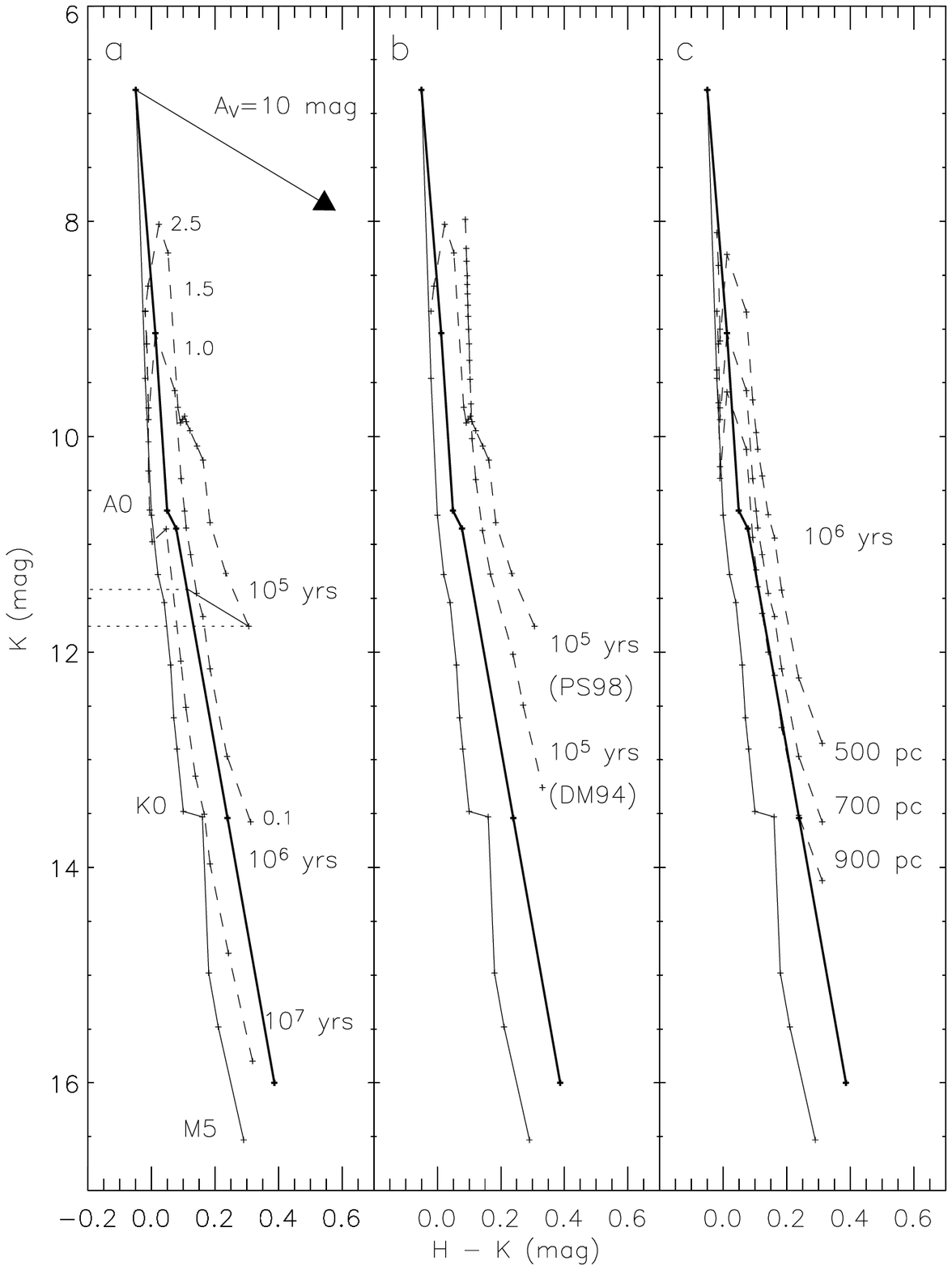}}
\caption[]{{\bf a} Colour-magnitude diagram showing the ZAMS locus
(thin solid line), three isochrones ($10^{5}$, $10^{6}$ and $10^{7}$ yr)
from the pms tracks by Palla \& Stahler (1999; dashed lines) and
a ``mean'' locus used to deredden NIR sources (thick solid line), all
for a distance of 700 pc. A few spectral
types on the ZAMS and star masses on the $10^{6}$ yr isochrone are
labelled, along with the ages of isochrones. An arrow shows the reddening
for 
$A_{V} = $ 10 
mag according to Rieke \& Lebofsky (1985). The error
made in dereddening a 0.1-M$_{\sun}$ $10^{5}$-yrs-old pms star is
indicated by the horizontal short-dashed lines. {\bf b} same as {\em a}, but
only the ZAMS locus, the $10^{5}$ yr isochrones from Palla \& Stahler
(1999; labelled as PS98) and D'Antona \& Mazzitelli (1994; labelled as
DM94) and the mean locus are drawn. {\bf c} same as {\em a}, but 
only the ZAMS locus, the $10^{6}$ yr isochrone from Palla \& Stahler
(at a distance  of 500, 700 and 900 pc, labelled)
and the mean locus are drawn.
\label{dere:tec}} 
\end{figure}

Their youth was also confirmed through a preliminary modelization, constructing
reddened theoretical KLFs based on a rough estimates  
of the extinction distribution in the clusters according to the CCDs. 
The comparison with the observed cluster KLFs 
indicates that the clusters' age roughly ranges between $\sim 10^{6}$ and $\sim 10^{7}$ yr. 
This suggested us a simple method in order to deredden the observed KLF, which is
illustrated in Fig.~\ref{dere:tec}. Figure~\ref{dere:tec}a details a CMD 
($K$ vs. $H-K$) including the ZAMS locus (solid line) and three isochrones from the
evolutionary pms tracks of Palla \& Stahler (1999; dashed lines), all for a
distance of 700 pc. Clearly, pms stars
from $10^{5}$ years old to the ZAMS span quite a limited range in colour (0.2--0.4 mag)
and, at the same time, the slope of the reddening law is small. This means that the stars
may be dereddened on a CMD simply by projecting them back along the
reddening vector onto a mean locus (thick line in Fig.~\ref{dere:tec}) 
chosen to be located between the
isochrones. The error is small; e.\ g., for pms stars older than $10^{6}$ yr, the error
in $K$ due to the procedure is $\la  0.2$ mag, less than the adopted KLF binning interval
(0.5 mag). Figure~\ref{dere:tec}a shows the worst case, one of a 0.1-M$_{\sun}$
$10^{5}$-yrs-old pms star, whose obtained $K$ is 0.28 mag less than
its actual $K$ after dereddening to the mean locus. Hence, only if a large
fraction of cluster members were very young low-mass stars, the dereddened KLF
could be significantly deformed. Even so, Fig.~\ref{dere:tec}b shows that
this result depends on the adopted evolutionary tracks: using those of
D'Antona \& Mazzitelli (1994) somewhat reduces the discrepancy. Finally,
Fig.~\ref{dere:tec}c shows that an error on the cluster distance cannot cause
a significant deformation of the dereddened KLF with respect to the actual
``unreddened'' one.

%
%
%

\begin{figure*}
\centerline{\includegraphics[width=12cm]{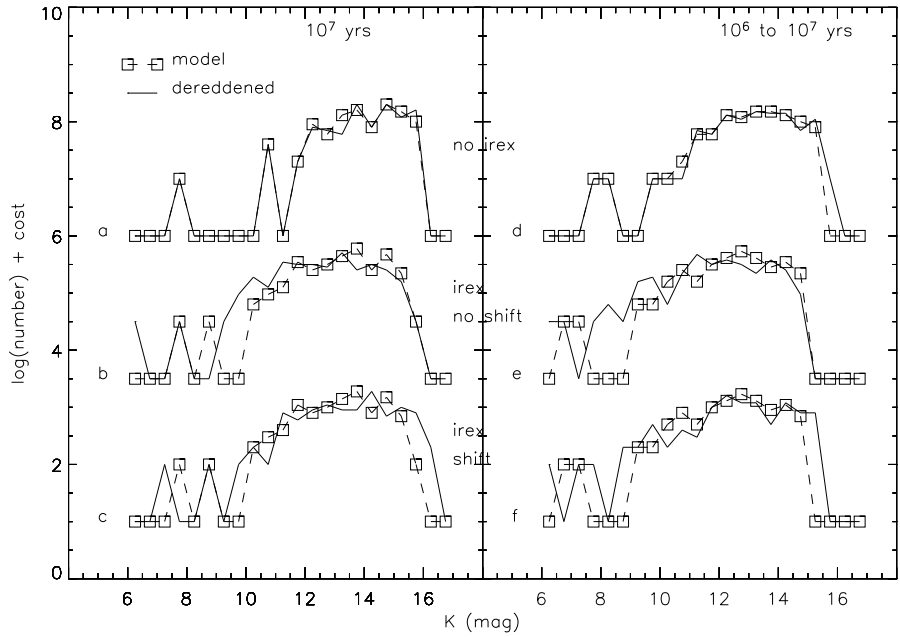}}
\caption[]{Comparison of model ``unreddened'' KLFs (open squares, dashed line), 
from one of the statistical realizations of a cluster at 700 pc
as explained in the text, and ``dereddened'' KLFs (solid line), obtained adding a random
extinction to each member of the cluster and then dereddening it according to 
our proposed method. Case {\em a} is for a coeval $10^{7}$ years old cluster without
NIR excess, case {\em b} includes a NIR excess and in case {\em c} the mean locus for dereddening
has been shifted by 0.4 mag redward in $H-K$ in order to partially compensate for the
NIR excess. Cases {\em d,e} and {\em f} same as {\em a, b} and {\em c}, 
but for a cluster modelled with a continuous
star formation from $10^{6}$ to $10^{7}$ years. Bins with a number of counts $N=0$ have
been set to $\log N =-1$ and a constant equal to 7,4.5 and 2.0 has been added to
$\log N$ for {\em ad, be} and {\em cf}, respectively.
\label{klf_c1:f}} 
\end{figure*}

How well the dereddening procedure works is illustrated in Fig.~\ref{klf_c1:f}.
A model ``unreddened'' KLF is derived from one of the statistical realizations
of a cluster with the IMF
of Scalo (1998), truncated at 0.1 M$_{\sun}$, 
and using the evolutionary tracks of Palla \& Stahler (1999).
For each star the $JHK$ magnitudes are scaled to a distance of 700 pc and 
reddened by adding a random value (ranging from 0 to 20 mag)
to an overall $A_{V}=20$ mag (screen), 
assuming the extinction law of Rieke \& Lebofsky (1985).
The synthetic sample is then projected onto our adopted mean locus in
the CMD and a ``dereddened'' KLF
is constructed. Cases a and d in Fig.~\ref{klf_c1:f} compare unreddened
and dereddened KLFs for a coeval cluster of 
$10^{7}$ years old and for one with continuous star formation from $10^{6}$ to $10^{7}$ 
years old, respectively, when no NIR excess is superposed to the photospheric emission. 
It is evident that the ``model'' and the ``dereddened'' KLFs are almost identical. 

Next, we tested how much the dereddened KLFs are affected by the NIR infrared
excess of the clusters' members.  Thus, we added
to each star of the synthetic sample a random $K$ contribution
modelled as in Hillenbrand \& Carpenter (2000). As shown in Fig.~\ref{klf_c1:f} for 
cases
b and e (same as a and d but with a NIR excess), the high luminosity end of the
``dereddened'' distribution stretches with respect to the ``unreddened'' KLF. 
A slightly better matching
is obtained by shifting the mean locus in $H-K$ redward of $0.4$ mag (roughly half
the maximum $H-K$ excess found by Hillenbrand \& Carpenter 2000), as illustrated by
cases c and f.

%
%
%

\begin{figure*}
\centerline{\includegraphics[width=12cm]{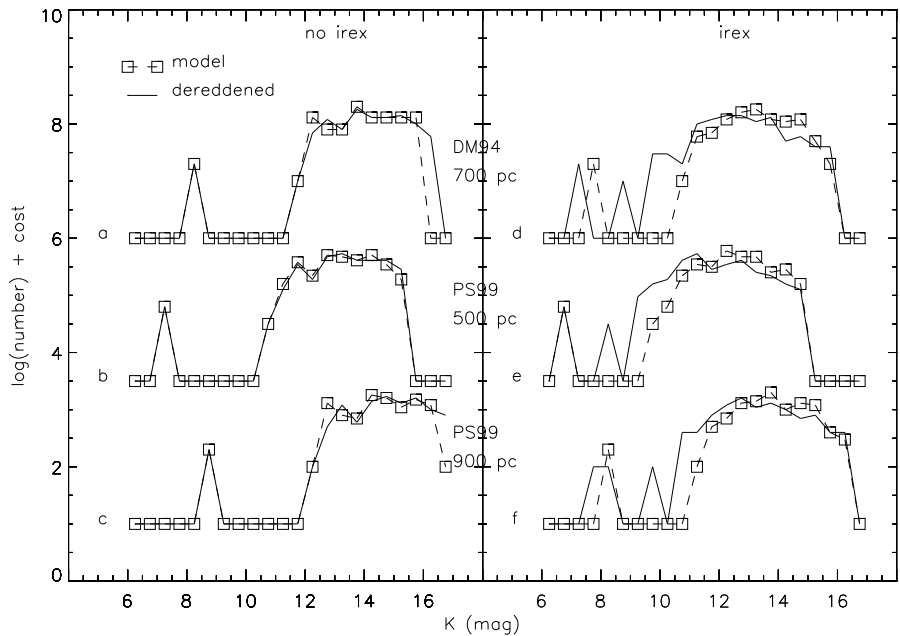}}
\caption[]{Comparison of model ``unreddened'' KLFs (open squares, dashed line), 
for the same cluster (of $10^{7}$ yrs) as case {\em a} 
of Fig.~\protect\ref{klf_c1:f}\protect,
and ``dereddened'' KLFs (solid line), obtained adding a random
extinction to each member of the cluster and then dereddening it according to 
our proposed method. 
Cases {\em a} and {\em d} are for the evolutionary tracks of D'Antona \& Mazzitelli
(1994) without ({\em a}) and with ({\em d}) a random NIR excess.
In cases {\em b} (without NIR excess) and {\em e} (with NIR excess) the cluster
has been put to a distance of 500 pc, whereas in cases {\em c} (without NIR excess)
and {\em f} (with NIR excess) it has been put to a distance of 900 pc.
Bins with a number of counts $N=0$ have
been set to $\log N =-1$ and a constant equal to 7.0, 4.5 and 2.0 has been added to
$\log N$ for {\em ad, be} and {\em cf}, respectively.
\label{klf_c2:f}} 
\end{figure*}

The effect of changing the distance of the synthetic cluster or the
evolutionary tracks is shown in Fig.~\ref{klf_c2:f}. The cluster of case a
above (coeval star formation, $10^{7}$ yrs old) 
was again synthesized but using the tracks of D'Antona \& Mazzitelli (1994), 
both without (case a) and with (case d) NIR excess; it is evident that the
choice of the evolutionary tracks underlying the modelled 
cluster does not affect much the dereddened
KLF which is obtained, whereas the effect of the NIR excess is the same
as in Fig.~\ref{klf_c1:f}.

The mean locus has been derived ``theoretically''
by assuming a distance of 700 pc, so we tested also what could happen if 
the actual clusters were at a different distance. Thus the cluster of
case a in Fig.~\ref{klf_c1:f} (coeval star formation, $10^{7}$ yrs old,
tracks of Palla \& Stahler 1999) was put at different distances.
In Fig.~\ref{klf_c2:f} the dereddened KLF is shown for
a distance of 500 pc (b without NIR excess, 
e with NIR excess) and 900 pc (c without NIR excess, f with NIR excess).
Clearly,
the ``dereddened'' KLF always reproduces quite well the ``unreddened'' KLF when no NIR
excess is included. When a random NIR excess is added, the deformation of the 
``dereddened'' KLF at the high luminosity end does appear similar 
for all cluster models illustrated in Figs.~\ref{klf_c1:f} and \ref{klf_c2:f}. 

We repeated this kind of visual test for a number of different synthesized clusters,
not shown here, varying the number of members, the age and the star formation history. 
We found that when no NIR excess is included
the ``dereddened'' KLF always matches quite well the ``model'' one for 
coeval clusters of $10^{6}$ yrs old, whereas behaves 
slightly worse for $10^{8}$ yrs old clusters. Anyway, the surveyed clusters
are very unlikely to be so old.
The ``dereddened'' KLF for
extremely young clusters ($\la 10^{5}$ yrs old) should appear narrower than
the actual one, mimicking an even younger age; we did not perform experiments
on such an unevolved star population, since probably the obtained KLF is much more sensitive
to the adopted evolutionary model (see, e.\ g., Fig.~\ref{dere:tec}).

%
%
%

We carried out also another simple test in order to get an unbiased assessment
of the goodness of the dereddening technique. 
We
constructed 6 synthetic coeval clusters of age ranging from $10^{5.5}$ to $10^{8}$
years in steps of 0.5 (in logarithm). From these, we got the actual 
``unreddened'' KLFs with and without a random NIR excess component. 
In both cases, a variable reddening was added to each synthetic member
and the KLFs were then dereddened, using both the
original mean lous and the one shifted in $H-K$ by 0.4 mag to compensate for the NIR
excess. Each of the ``dereddened'' KLFs was compared to all the ``unreddened'' ones 
and the mean squared differences ($\chi^{2}$)
in counts per magnitude bin (of width 0.5 mag) were computed. 
The results can be summarized as follows:
\begin{enumerate}
\item when no NIR excess is included, the least $\chi^{2}$ is obtained 
for the same age;
\item when a NIR excess component is included, the best fit is generally 
obtained for the same 
age, but an ambiguity arises between the $10^{6.5}$ yrs old and the
$10^{7}$ yrs old models. The same occurs when shifting the mean locus in $H-K$,
although the obtained $\chi^{2}$ values are lower than before;
\item using the evolutionary tracks of D'Antona \& Mazzitelli (1994)
in synthesizing the clusters, the
best fits are always obtained for the same age, excepted  that the dereddened KLF
for a $10^{8}$ yrs old cluster (including a NIR excess component) is equally well fitted
by unreddened KLFs  $10^{7.5}$ and $10^{8}$ yrs old; 
\item when the cluster distance is set to 500 pc, the best fits are for the
same age, but when NIR excess is included, the $10^{6}$ and $10^{7}$ yrs old
``dereddened'' KLFs look younger by 0.5 (in logarithm). Shifting the mean
locus in $H-K$ corrects for the effect, but causes the $10^{7.5}$ yrs old
``dereddened'' KLF to be fitted by the $10^{8}$ yrs old KLF;
\item when the cluster distance is set to 900 pc, the best fits are always for
the same age, excepted that shifting the mean locus in $H-K$ causes the $10^{6.5}$
yrs old ``dereddened'' KLF to be well fitted by the $10^{7}$ yrs old KLF.
\end{enumerate}

We performed similar tests by constructing 3 new synthetic clusters with
continuous star formation in the ranges $10^{5.5}-10^{6.5}$, $10^{6.5}-10^{7.5}$ 
and $10^{7.5}-10^{8.5}$ yrs. Now, the ``dereddened'' KLF is always best fitted
by the ``unreddened'' KLF for the same cluster age range, unrespective of the presence
of a random NIR
excess. Comparing the ``dereddened'' coeval KLFs to the 3 unreddened ones 
with continuous star formation, indicates that
only for the oldest coeval clusters some ambiguity may arise with KLFs of 
evolved cluster with continuous star formation. In conclusion, our dereddening
technique allows to reconstruct the actual KLF with sufficient accuracy provided
that the adopted mass-luminosity relation and distance to the VMR are
within reasonable values and that 
no NIR excess is included. 
When the latter is not negligible,
the dereddening technique stretches the ``dereddened'' KLF at the high luminosity end
with respect to the ``unreddened'' one, but the age of the cluster appears still 
discernible within a factor of $10^{0.5}$.

\begin{figure}
\centerline{\includegraphics[width=6cm]{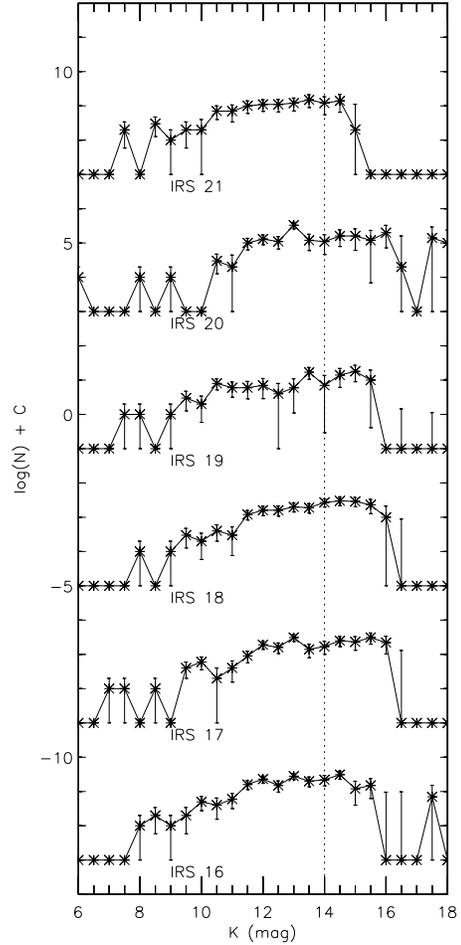}}
\caption[]{The KLFs of the 6 clusters, corrected for the field stars
contribution and dereddened according to the method described in the
text. The vertical dotted line marks the estimated completeness limit.
\label{klf_ld:f}} 
\end{figure}

By dereddening the KLFs of the observed clusters as explained,
we found only small differences between the dereddened KLFs 
either using the mean locus or the shifted one. So,
at last we decided to deredden the clusters' KLFs onto the unshifted mean locus.
This also because, as can be seen in Fig.~\ref{col-col-in}, only a fraction of
sources exhibit a NIR excess, whereas our tests always assumed that {\em all}
sources had a NIR excess component. Furthermore, it is likely that the most prominent
NIR excess is displayed by the brightest $K_{s}$ sources, i.\ e., those 
falling in a luminosity interval where the KLFs are 
anyway scarcely sampled due to the poor statistic. Thus, the dereddened KLFs
should be well representative of the actual unreddened ones.
These are shown in Fig.~\ref{klf_ld:f}. It is
evident that all are quite similar, both in extent and in shape. An estimate
of the completeness limit has been computed as follows. A vertical line
is drawn enclosing the maximum $H-K$ displayed by the sources 
in the CMDs and its intersection is found with the
line of the photometric completeness limit. Projecting this point along the 
reddening vector onto the mean locus then yields the ``dereddened'' completeness 
limit. This is estimated to be $K_{s} \sim 14$ and it is shown in
Fig.~\ref{klf_ld:f} as a vertical dotted line.

All KLFs appear extremely flat across the ``dereddened''
completeness limit, which corresponds to pms stars of $\sim 0.1$ M$_{\odot}$
for an age of $10^{6}$ yrs ($\sim 0.4$ M$_{\odot}$ for an age of $6 \times 10^{6}$ yrs),
according to the tracks of Palla \& Stahler (1999). This indicates that the
underlying IMF is flat or slightly increasing across this mass. A break on the
slope of the KLFs can be roughly identified at $K_{\rm s} > 10$, suggesting
a break in the IMFs at $< 2.5$ M$_{\sun}$ for $10^{6}$ yrs ($<3.5$ M$_{\sun}$ for
$6 \times 10^{6}$ yrs). Above this mass limit, the source statistic is quite poor.
The magnitude shifts of the KLFs with respect to each other due to small
differences in the clusters' distance can be estimated as follows. The maximum
projected distance between clusters is $\sim 1.5\degr$, corresponding to $\sim 18$
pc. Hence, the maximum distance difference along the line of sight is expected
of the same order. This would cause maximum variations in the distance modulus
$\sim 0.06$ mag, much less than the binning width. 

The hypothesis that all KLFs arise from a same distribution can be tested
now. If plausible, this would confirm that both the underlying IMF and the
star formation history can have been similar for all clusters. Hence, we performed the
chi-square test on all pairs of KLFs; the obtained significance levels
are listed in Tab.~\ref{chi:2}. The KLFs are compared in the range
$K_{s} = 10 - 14$; the lower limit is chosen to avoid counting problems due
to saturation of the array. It can be seen that the hypothesis cannot be
disproved at generally high significance levels. IRS 16, 17 and 18 exhibit
quite high probabilities to represent a same distribution, whereas IRS 19 and 20  
are less likely to meet the hypothesis, but still this cannot be disproved.
Performing the test on the dereddened KLFs uncorrected for the field
contribution generally causes slightly higher levels for the larger significance
cases and lower levels for the smaller significance cases, but of the same
order as before.
In summary, the probability that the observed KLFs are drawn from a same
distribution appears significant.
%
\begin{table}
\caption[]{Significance level of the hypothesis that two cluster KLFs
 are the same distribution from the chi-square test. The crossing
 of each line and column lists the significance level for that pair
 of KLFs.
\label{chi:2}}
\begin{tabular}{lllllll}\hline
Field & IRS16& IRS17 & IRS18 & IRS19 & IRS20 & IRS21\\ 
name  &        &      &        &          &        &       \\
\hline
IRS16 &      & $0.95$ & $0.93$ & $0.26$ & $0.32$ & $0.79$ \\ 
IRS17 & $0.95$ &      & $0.61$ & $0.06$ & $0.52$ & $0.25$ \\ 
IRS18 & $0.93$ & $0.61$ &      & $0.20$ & $0.34$ & $0.71$ \\  
IRS19 & $0.26$ & $0.06$ & $0.20$ &      & $0.04$ & $0.96$ \\ 
IRS20 & $0.32$ & $0.52$ & $0.34$ & $0.04$ &      & $0.28$ \\ 
IRS21 & $0.79$ & $0.25$ & $0.71$ & $0.96$ & $0.28$ &      \\ 
\hline
\end{tabular}
\end{table}
%
%
%

%
%
%

\subsection{Deriving the initial mass function of the clusters}
\label{dimfc}

As a last step, we want to derive the IMF underlying the clusters' KLFs.
So far, most of the efforts in the literature have been devoted towards
constructing synthetic KLFs from a given IMF and star formation history
(see, e.\ g., Lada \& Lada 2003). These are then fitted to the observed KLFs and the 
IMF underlying the best-fit KLF is thus extracted. 
Due to the limited statistic of our clusters,
we chose to directly obtain the IMF by using the relation:
\begin{equation}
\label{klf:imf}
\frac{{\rm d}N}{{\rm d} \log M} = \frac{{\rm d}N}{{\rm d} K} 
(\frac{{\rm d} \log M}{{\rm d} K})^{-1}.
\end{equation}
Note that this IMF is still model-dependent since the mass-magnitude 
relation has to be derived from a set of evolutionary tracks. And a star
formation history has to be assumed.

\begin{figure}
\centerline{\includegraphics[width=8cm]{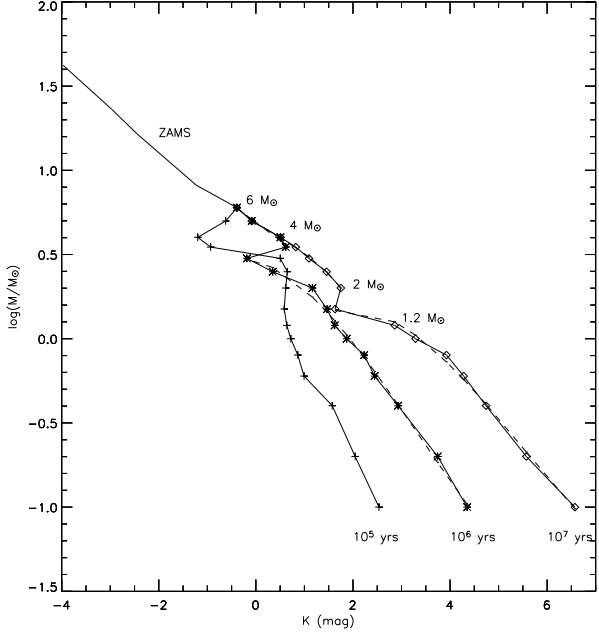}}
\caption[]{Mass-magnitude relation for pms stars $10^{5}$ (plusses),
$10^{6}$ (asterisks) and $10^{7}$ (open diamonds) yrs old, from
the evolutionary tracks of Palla \& Stahler (1999). Masses range
from 6 (upper end) to 0.1 (lower end) M$_{\sun}$ (a few points are
indicated). Fits to the tracks are shown as dashed lines.
The mass-magnitude relation for ZAMS stars is also drawn ($M > 6$
M$_{\sun}$.).
\label{mass_mag}} 
\end{figure}

We computed the mass-magnitude relation from the evolutionary tracks of Palla \&
Stahler (1999) for a grid of ages spanning $10^{5}$ to $10^{7}$ yrs. 
If the clusters were coeval, Eq.~\ref{klf:imf} could be applied straightforwardly
and the IMF could be obtained directly from the KLF. This because 
the freezing of time
allows one to fix the mass-magnitude relation and
a given magnitude corresponds uniquely to a given mass (excepted in small magnitude
intervals; see below).
However, in order to apply
Eq.~\ref{klf:imf} to the clusters' KLFs, both the KLFs and the derivative 
of the mass-magnitude
relation have also to change slowly. 
Whereas this is true for the KLFs, 
it can be seen in Fig.~\ref{mass_mag} that there
are limited intervals in magnitude where this condition is not satisfied by the
derivative of the mass-magnitude relation. In this respect,
there appear to be two kinds of
problems. The first one is evident for very young ages (see the track $10^{5}$ yrs old), 
where ${\rm d} \log M/{\rm d} K \rightarrow \infty$ around $1.5 M_{\sun}$. 
This is a signature of the birthline (see, e.\ g., Fig~10 of Palla \& Stahler
1993) and, hence, reflects the initial conditions of the pms contraction.
The other problem is common to all ages and is represented by a sort of ``tooth''
moving from higher to lower masses at increasing ages. This causes a small magnitude
interval to be ``degenerate'' in mass (i. e., stars with different masses
display the same $K$). This appears to be related to the end of the radiative
contraction and the start of the main sequence.
We avoided an interval spanning 1 magnitude around the points where
${\rm d} \log M/{\rm d} K \rightarrow \infty$ and the whole range 
(increased by 0.25 mag at each end) where the masses are ``degenerate''. 
Outside these intervals, we used polynomial
fits to approximate the mass-magnitude relation (see Fig.~\ref{mass_mag}). In turn,
these have been used in order to convert $K$ into masses and to estimate
${\rm d} \log M/{\rm d} K$.

As for massive stars, according to the Palla \& Stahler scenario an object
enters the ZAMS after accreting 8-10 M$_{\sun}$.
Hence, we used the mass-luminosity relation as derived from the ZAMS for stars 
with $M > 6$ M$_{\sun}$. We adopted the absolute magnitudes given by Panagia
(1973) and the colours by Koornneef (1983) to obtain the $K$ magnitudes
of OB stars. The masses were then determined by converting the bolometric luminosities
given by Panagia (1973) according to the formulae by Tout et al.\ (1996).
This is also shown in Fig.~\ref{mass_mag}. For an assessment of the errors affecting
this relation see, e.\ g., Vacca et al.\ (1996).
%
%
%

\begin{figure*}
\centerline{\includegraphics[width=16cm]{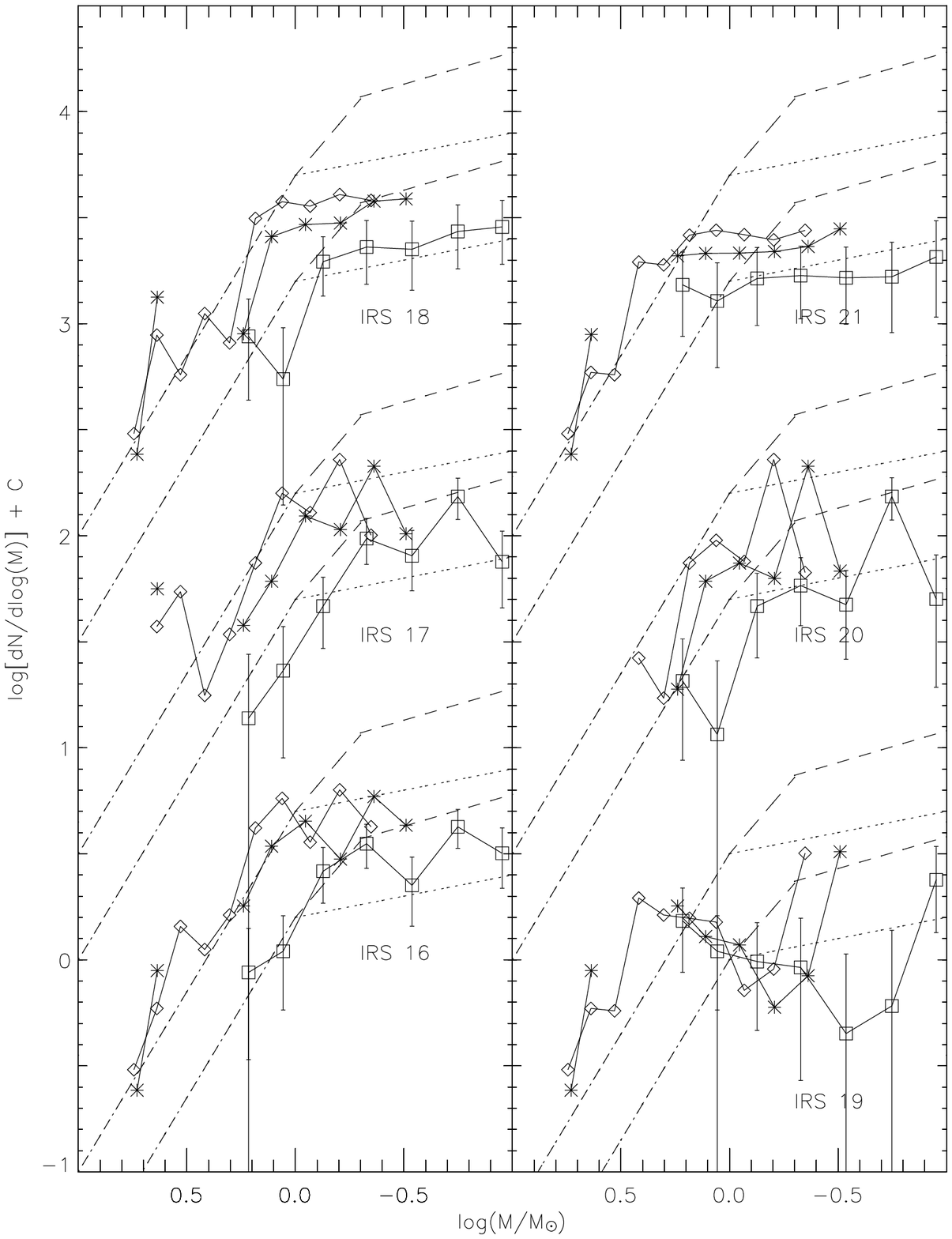}}
\caption[]{The initial mass functions for the 6 clusters, derived
assuming coeval star formation and an age of $10^{6}$ (open squares),
$3 \times 10^{6}$ (asterisks) and $6 \times 10^{6}$ (open diamonds)
yrs. All cluster IMFs are compared with the 
IMFs from Scalo (1998; dotted
line) and Kroupa et al.\ (1993; dashed lines), each with two different
normalizations. For the sake of clarity, errorbars are drawn only for
$10^{6}$ yrs old.
\label{cl:imf}} 
\end{figure*}

The clusters' IMFs for assumed ages of $10^{6}$, $3\times 10^{6}$ and
$6\times 10^{6}$ yrs and coeval star formation, are shown in Fig.~\ref{cl:imf}.
These are compared with the IMFs from Scalo (1998) and Kroupa et al.\ (1993).
The Scalo's IMF (in the form ${\rm d}N/{\rm d}\log M$) is approximated by
a power law with $\Gamma = -1.7$ between 10--1 M$_{\sun}$ and
$\Gamma = -0.2$ between 1--$0.1$ M$_{\sun}$. It is derived by a
compilation of different determinations.  That of Kroupa et al.\ (1993)
is representative of field stars and is similar to the Scalo's (1998).
In particular, $\Gamma = -1.7$ for $M >1$ M$_{\sun}$, $\Gamma = -1.2$
between 1--$0.5$ M$_{\sun}$ and $\Gamma = -0.3$ between $0.5$--$0.08$ 
M$_{\sun}$.

The errors, propagated from the statistical uncertainties of the KLFs, are drawn only
for $10^{6}$ yrs old.
Because of the intervals to be avoided, the IMF $10^{6}$ yrs old is not defined
for $M > 1.6$ M$_{\sun}$ whereas the one $3 \times 10^{6}$ yrs old exhibits a gap
between $1.6$ and $4$ M$_{\sun}$. 
Moreover, at this stage we do not show the IMF for the most massive stars 
through 
the ZAMS mass-luminosity relation because of the poor statistic. 
However, it is clear that the clusters' IMFs 
are consistent
with the ``comparison'' ones
within the uncertainties. In the case of IRS 19,
the shape of the IMF seems to be affected by the poor statistic rather than by 
actual differences 
with the comparison IMFs.

In a few cases, a better agreement would be
obtained by shifting the break point envisaged in the Scalo (1998) IMF
from 1 to $\sim 1.5$ M$_{\sun}$, 
especially if older ages are assumed. 
But it is difficult to assess whether this
reflects an actual difference.
We also checked that using the evolutionary tracks of D'Antona \&
Mazzitelli (1994) does not affect significantly the obtained IMFs,
excepted for $10^{5}$ yrs.

Assuming a coeval star formation is obviously an oversimplification.
Nevertheless, if star formation is peaked at a given age, this
should not change the results. Furthermore, it is evident from
Fig.~\ref{mass_mag} that the derivative of the mass-magnitude relation
changes slowly with time between $1-2$ and $0.1$ M$_{\sun}$. What 
changes is the $K$ magnitude. We can hope that the results should be
still little affected even in the case of continuous star formation if 
it lasts for a time such that the
change in $K$ is $\la 0.5$ mag (i.\ e., the binning interval of our
KLFs). This probably translates, e.\ g.,
into a duration of $\la 2 \times 10^{6}$ yrs 
for ages of a few $10^{6}$ yrs.  
Some concerns arose from the possibility that star formation may
span $\sim10^{7}$ yrs and accelerate, as proposed by Palla \& Stahler
(1999). Hence, we approximated clusters of $\sim 100-200$ members
with accelerated star formation
by adding a KLF synthesized in the coeval $10^{6}$ yrs old case to a KLF 
synthesized in the case of continuum star formation from $10^{7}$ to $10^{6}$ yrs,
with different fractions of the total member number. Then, we derived the
IMF from the cumulative KLF by using our method. We checked that the
obtained IMF is very similar to the underlying one (Scalo-like)
in the $1.5$--$0.1$ M$_{\sun}$ range
even if up to a 35 \% of the synthesized population is produced through
continuum star formation. In particular, in a
$\log ({\rm d}N/{\rm d} \log M)$ vs. $\log M$ plot the purely
coeval IMF and the mixed one still coincide within the statistical errors.
This may be understood analytically: the continuum component of the
KLF is flatter and spread on a larger magnitude interval than the
coeval component, and its underlying area is also much smaller.
But we checked that even increasing the continuum fraction of the
synthesized population to a 50 \%, the obtained IMF does not change
dramatically, although appearing slightly flatter. However, this assumes
that the underlying IMF is the same in the two population components;
only in the unlikely case that stars of different masses are originated
at very different times
we could expect major differences between the obtained
and the actual IMF in the $1.5$--$0.1$ M$_{\sun}$ range.

We could reverse our point by trying to constrain the clusters' 
age assuming that the underlying IMF is 
a standard one, like our comparison IMFs.
If so,
the clusters' KLFs are consistent with coeval star formation and an age between
$1-6 \times 10^{6}$ yrs. Younger ages can be discarded since they
produce IMFs peaked at $0.3-0.1$ M$_{\sun}$ with Salpeter-like slopes.
Also, clusters older than $10^{7}$ yrs can be discarded, since the resulting
IMFs would be peaked at $> 2$ M$_{\sun}$ and declining 
towards higher and smaller masses.

%
%
%

\section{Discussion}
\label{disc:s}

Knowing the dereddened cluster KLFs allows us to estimate the number
of members $N_{\rm T}$ up to the completeness magnitude 
(i.\ e., down to $0.1$--$0.4$
M$_{\odot}$), to be compared with I$_{\rm c}$ in Table~\ref{clust:prop}.
The results are listed in Table~\ref{last:par} and 
it can be noted that $N_{\rm T}$ is roughly twice
the corresponding value of I$_{\rm c}$. Clearly, the latter underestimates
the true number of members, probably due to the fact that the sky contribution
is estimated in a region still contaminated by stars from the cluster population. 
In fact, deriving the clusters' KLFs within a radius of 1 arcmin (instead
of 2 arcmin) results in a much lower $N_{\rm T}$.

If we want to include in $N_{\rm T}$ all clusters' members down to the hydrogen
burning limit, then the values derived from our observations have
to be corrected for the age.
Having proved that the clusters' IMF is very similar to 
standard ones, like those of Scalo (1998) or Kroupa et al.\ (1993),
by adopting for it the IMF of Scalo (1998) we can estimate the 
correction to apply to $N_{\rm T}$. 
E.\ g., for a $6 \times 10^{6}$ yrs old
cluster, our completeness magnitude corresponds to a mass limit of $0.4$
M$_{\sun}$. Integrating the Scalo IMF from $0.4$ to $0.1$ M$_{\sun}$
yields $\sim 55$ \% of stars in that mass range, so the observed $N_{\rm T}$ 
should be increased by a factor of $2.2$.
This largely exceeds the statistical (counting) uncertainties listed in
Table~\ref{clust:prop}. Note that a further underestimate is likely to arise from
having selected sky areas close to the clusters in order to construct the
control KLFs (i.\ e., possibly still containing a small number of clusters' members).
And we do not consider here the brown dwarf population
(which is not sampled by our observation). 

Actually, $N_{\rm T}$ is bound to be underestimated anyway, due to the
unknown fraction of close companions which escape observations.
How critical this problem may be is demonstrated by Fig.~6 of
Giannini et al.\ (2005); using ISAAC images of IRS 17, these authors can resolve
source \# 40 of Massi et al.\ (1999) into 6 different components. We checked
that the same star in our $K_{s}$ image is resolved into only 3 components,
although the brightest ones.

The total mass (down to the hydrogen burning limit) of the clusters can 
be derived, as well, by integrating the Scalo IMF times the mass. The results are listed
in Table~\ref{last:par} and, like $N_{\rm T}$, can be underestimated
by a factor 2.2 or so, depending on the age. The highest expected
stellar mass $x$ in each cluster was estimated
by integrating the Scalo IMF, as well, from $\infty$ to a mass $x$ and requiring  
that the result is 1. The obtained values (see Table~\ref{last:par}) are
consistent with the bolometric luminosities given by Massi et al.\ (2000, 2003)
and confirm that the most massive stars in the 6 clusters are intermediate-mass
stars. Note that an increase in $N_{\rm T}$ of 
a factor $2.2$ 
would roughly double $x$.

At last, we can derive the star formation efficiencies of the parental clumps
if the mass of the associated gas is known. We used the mm (C$^{18}$O) observations of
Wouterloot \& Brand (1999) and Yamaguchi et al. (1999). Because of the low
resolution of the latter data, these probably refer to matter on a larger scale
than found by Wouterloot \& Brand (1999), i.\ e.\ clouds
rather than clumps or cores, and are likely to underestimate the gas content of
these larger scale structures. Nevertheless, all derived efficiencies are 
$\leq 26$ \%,
excepted towards IRS 20 where a star formation efficiency of $\sim 80$ \%
is found. These
are typical values for young embedded clusters (Lada \& Lada 2003). 
Instead, if taken at face value, the high star formation efficiency towards IRS 20
implies that the cluster is emerging as a bound structure (Lada \& Lada
2003); 
however, we remark that the area around IRS 20 covered by the mm observations of 
Wouterloot \& Brand (1999) is very limited and the gas mass could then have been
underestimated.

Although our NIR observations were able to clarify the {\em shape} of
the clusters' IMF between 1--3 and $0.1$--$0.4$ M$_{\sun}$, they fail to
strongly constrain the clusters' age. This would have allowed to more accurately
model the dereddened KLFs and derive the underlying IMF. Also, the
substellar mass regime is not probed by our observations, requiring much
deeper imaging. 
However, as shown in Fig.~\ref{klf_ld:f}, the dereddened KLFs  
maintain roughly the same slope above the completeness magnitude
(remember that 5 fields are $\sim 1$ mag deeper than the 6th one). 
No strong decrease is evidenced; hence, because of uncompleteness,
the KLFs can at most increase above this limit. This
means either that the clusters are slightly older than $1 \times 10^{6}$
yrs and stars fainter than $K_{s} = 14$ are still above the hydrogen burning
limit (and follow a Scalo-like IMF), or that the clusters are
as old as, or younger than, $1 \times 10^{6}$ yrs and the IMF itself is still
increasing at the hydrogen burning limit (assumed ${\rm d} \log M/{\rm d}
K$ does not exhibit major changes below $0.1$ M$_{\odot}$).

As for the high mass end of the IMF, this is not only
affected by the poor statistic. The brightest sources in our $JHK_{\rm s}$
images are also saturated and, probably, most of the brighter $K_{\rm s}$
sources are largely affected by a NIR excess emission. However, the
data obtained by Massi et al.\ (1999, 2003) are less
affected by saturation or non-linearity problems. As for IRS 16, it coincides
with an HII region and these authors found that it is excited by a B0--B5
source, which is the most massive cluster's member 
(consistently
with the highest expected stellar mass listed in Table~\ref{last:par},
given also the uncertainty on $N_{\rm T}$).
In the case of the remaining clusters, the most massive clusters'
members are the sources emerging in the NIR and identified by
Massi et al.\ (1999) as the counterparts of the IRAS point sources. 
Whereas the clusters' IMF is consistent with the mass of the most massive
star in each cluster, imaging surveys of IRAS point sources in the VMR
have so far failed to unveil high luminosity IRAS sources associated with
isolated young stars (Massi et al.\ 1999, 2003). 

%
\begin{table}
\caption[]{Clusters' properties derived from their IMF. 
Column 2 lists the number of cluster members
(down to the mass completeness limit), Col.\ 3 the cluster
total mass (as in Col. 2), Col.\ 4 the mass of the most massive star
within the cluster and Col.\ 5 the star formation 
efficiency. \label{last:par}}
\begin{tabular}{ccccc}\hline
Field & Members & Mass & Largest & SFE \\ 
name  &        &      & star & \\
      &        & (M$_{\sun}$) & (M$_{\sun}$) &              \\
\hline
IRS 16 & $144 \pm 15$ & 103 & $7.1$ & 0.24$^{a}$ \\
IRS 17 & $126 \pm 15$ &  90 & $6.5$ & 0.26$^{b}$ \\
IRS 18 & $124 \pm 14$ &  89 & $6.5$ & - \\
IRS 19 & $69 \pm 13$  & 49.5 & $4.4$ & 0.095$^{b}$ \\
       &              &      &   & 0.21$^{a}$ \\    
IRS 20 & $98 \pm 14$  &  70 & $5.5$ & 0.8$^{b}$ \\
IRS 21 & $95 \pm 14$  &  68 & $5.4$ & 0.11$^{a}$\\
\hline
\end{tabular}

\vspace*{1mm}
$^a$~gas mass from Yamaguchi et al.\ (1999) \\
$^b$~gas mass from Wouterloot \& Brand (1999)

\end{table}
%
%
%

\begin{figure}
\centerline{\includegraphics[width=8cm]{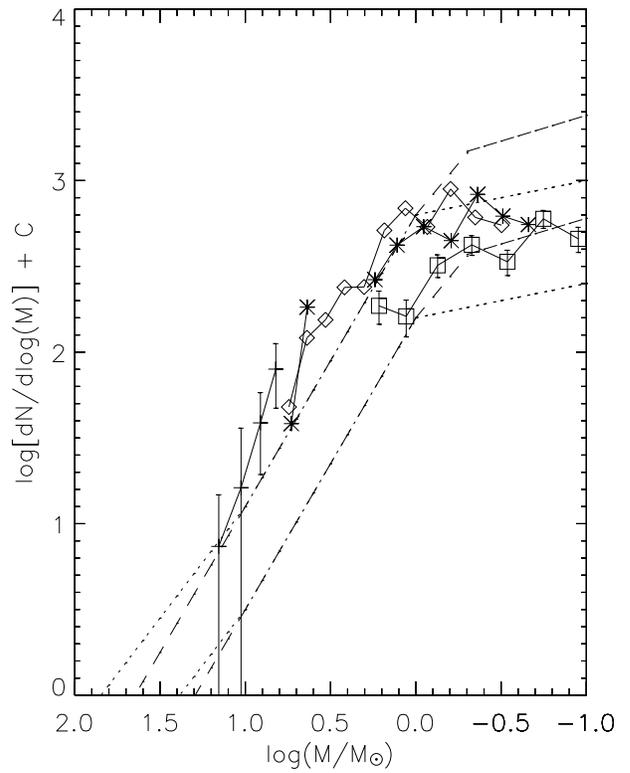}}
\caption[]{Sum of the 6 initial mass functions, 
same symbols as Fig.~\protect\ref{cl:imf}\protect.
The four most massive bins have been derived using the
ZAMS mass-magnitude relation discussed in the text.
\label{total:imf}} 
\end{figure}

A way to investigate the high mass end of the clusters' IMF is by
adding together all 6 IMFs, as to increase the statistic. This should also 
approximate the field star IMF, if most stars originate in clusters and then
disperse. The ``global'' IMF is shown in Fig.~\ref{total:imf}. The four most massive
bins have been obtained by using the ZAMS mass-magnitude relation discussed in
Sect.~\ref{dimfc}. We dropped three points with resulting masses 
$> 20$ M$_{\sun}$. We checked that these are due to the young stars or protostars 
lying in the centre
of IRS 17, IRS 19 and IRS 20 and identified by Massi et al.\ (1999), then
members of the clusters. But, as shown by Massi et al.\ (2000, 2003), based
on the measured bolometric luminosity and the upper limit on the radio continuum
emission, they are intermediate-mass, not high-mass (proto-)stars. 
Clearly, they exhibit a strong 
NIR excess emission and the ZAMS mass-magnitude relation cannot be
applied to them.

The global IMF is quite similar to the comparison ones, particularly
at $> 10^{6}$ yrs. The IMF $10^{6}$ yrs old implies a break at 
$M > 1$ M$_{\sun}$, but the mass of many objects in the intermediate-mass
regime may have been overestimated because of the NIR excess.
At the high-mass end, the IMF slope appears slightly steeper than that of
the comparison ones, but here the statistic is still poor. However,
it is noteworthy a drop-off of the IMF at $M \sim 10$ M$_{\sun}$. 
There are 13 stars in the 4 most massive bins ($M \sog 6$ M$_{\sun}$),
of which 1 in the most massive one ($M \sim 14$ M$_{\sun}$) and 2
in the next most massive one ($M \sim 10$ M$_{\sun}$).
This indicates that none of the 6 clusters could produce a high-mass stars,
although they all originated intermediate-mass stars as predicted, e.\
g., by the Scalo's IMF based on their number of members. But applying
the same statistical argument, since we are now sampling a
total number of stars $> 656$
(see Table~\ref{last:par}) we should find a star with $M > 22.5$ M$_{\sun}$
(i.\ e., an O8--O9 ZAMS star).
Such an object would have a bolometric luminosity $> 10^{4}$ L$_{\sun}$
and form an HII region. There is not such a massive source within the 6 clusters
(Massi et al.\ 2000, 2003). Hence, the drop-off appears to be a real
feature of the global IMF.

If the actual clusters' IMF is a standard one, then the most massive stars
in each cluster depend on the cluster mass, as expected based on statistical
grounds. But they fail to follow statistics when added together, suggesting
that the dependence on the cluster mass is ``physical''. I.\ e., high-mass
stars need large clusters to be formed.
On the other
hand, it may be that the IMF in the whole region is a standard one, while that
of the single clusters is much steeper at the high mass end. Then, we should find
massive stars elsewhere in the region. Massi et al.\ (1999, 2003) do not
find massive stars in their D-cloud sample of IRAS sources, but those objects 
might well have escaped 
the selection criteria adopted by Liseau et al.\ (1991). This would
reconcile with a statistical view of the IMF, but if the clusters' IMF
is steeper than that of field stars or open clusters, it is likely so
because of different physical conditions in the parental gas, so, in a sense,
still ``physical''. A careful multi-wavelength examination 
of the whole region is then necessary in order to settle this point.
%
%
%

\section{Conclusions}
\label{conc:s}

We have presented the results of a Near-Infrared photometric survey of a sample
of 6 embedded young clusters, all associated with
luminous IRAS sources in the Vela D molecular cloud. 
They lie at the same distance and have been imaged with the
same instrumental setting;
this allowed us to derive the clusters' properties in
a consistent way, without most of the observational biases often affecting
the comparison of different star forming regions.
The clusters' $K$ Luminosity Functions have been obtained from the same
images, by subtracting
a control $K$ Luminosity Function, for sources $> 2\arcmin$ from the
clusters' centre, from that for sources $< 2\arcmin$ from the clusters' centre.
A simple method for dereddening the $K$ Luminosity Functions, based
on colour-magnitude diagrams, has been developed and described. 
The main results are the following:
\begin{enumerate}
\item Our images confirm the previous results of Massi et al.\ (2000, 2003),
 i.\ e., all selected fields contain young embedded clusters with a
 radius $\sim 0.1-0.2$  pc and a mean stellar density $\sim 10^{3}$ 
 stars pc$^{-3}$. We also resolved, for the first time, IRS 17 into 
 2 sub-clusters.
\item the dereddened clusters' $K$ Luminosity Functions have been compared
 in pairs using the chi-square test; the hypothesis that all arise from a same
 distribution cannot be disproved at significance levels ranging from
 $0.04$ to $0.95$ (but $> 0.50$ in most cases). This suggests that the
 underlying Initial Mass Function and star formation history may have been
 quite similar.
\item The clusters' Initial Mass Functions, obtained from the dereddened
 $K$ Luminosity Functions in the hypothesis of coeval star formation,
 are compatible with 
 standard ones, like those of Scalo (1998) and Kroupa et al.\ (1993),
 for assumed ages ranging from
 $10^{6}$ to $6\times10^{6}$ yrs. Major departures would occur if the clusters
 were either younger or older than assumed. However, they are unlikely
 to be older than $5 \times 10^{6}$ yrs since all of them are still associated
 with molecular gas.
\item The global IMF, obtained by adding together all 6 single IMFs,
 is similar to a standard IMF. But it appears steeper at the high-mass
 end and exhibits a fall off at $\sim 10$ M$_{\sun}$. E.\ g., on 
 statistical grounds the Scalo
 IMF would predict a star of $>22.5$ M$_{\sun}$ within one of the cluster,
 which is not found. Then, either high-mass stars need larger clusters
 to be formed, or the single IMFs are steeper at the high-mass end
 because of the physical conditions in the parental gas.
\item Based on their $K$ Luminosity Functions, the clusters' members range
 from 69 (IRS 19) to 144 (IRS 16). Assuming an Initial Mass Function like
 Scalo (1998) yields total clusters' masses ranging from 50 to 100 M$_{\odot}$.
 If the clusters are $10^{6}$ yrs old, the stellar mass is sampled down to
 0.1 M$_{\sun}$. If the clusters are as old as $6\times10^{6}$ yrs old, both
 number of members and mass have to be increased by a factor 
 $\sim 2.2$ to reach the same mass lower limit.
\item The star formation efficiency ranges from 0.1 to 0.8, being $< 0.26$ 
 in all cases but one (IRS 20). 
\end{enumerate}

\begin{acknowledgements}
We thank the ESO staff for the excellent support during 
the observations at the NTT.
We also thank the referee, Bruce Elmegreen, whose comments greatly improved
the quality of this work.
FM acknowledges partial support from ``fondi finalizzati INAF, bando 2002:
Disks and Jets in High-mass Star Forming Regions''.
\end{acknowledgements}
%
%

\end{document}